\documentclass[preprint,3p,12pt]{elsarticle}

\usepackage{amsmath,amssymb,amsfonts,bm,mathtools}
\usepackage{graphicx}
\usepackage{booktabs,multirow}
\usepackage{algorithm}
\usepackage[noend]{algpseudocode}
\usepackage{float}
\usepackage{flafter}
\usepackage{placeins}
\usepackage{xcolor}
\usepackage{hyperref}
\usepackage{amsthm}

\newtheorem{proposition}{Proposition}

\newtheorem{remark}{Remark}

\newcommand{\dd}{\mathrm{d}}
\newcommand{\vecomega}{\boldsymbol{\Omega}}

\newcommand{\vgrad}{\boldsymbol{\nabla}}

\begin{document}

\begin{frontmatter}

\title{A nonequilibrium-capturing unified gas-kinetic scheme II: multi-frequency radiation transport}

\author[HKUST1]{Hongyu Liu}
\ead{hliudv@connect.ust.hk}

\author[HKUST1]{Xiaojian Yang}
\ead{xyangbm@connect.ust.hk}

\author[XJTU]{Xing Ji}
\ead{jixing@xjtu.edu.cn}

\author[IAPCM]{Chang Liu}
\ead{liuchang@iapcm.ac.cn}

\author[HKUST1,HKUST2,HKUST3]{Kun Xu\corref{cor1}}
\ead{makxu@ust.hk}

\cortext[cor1]{Corresponding author}

\address[HKUST1]{Department of Mathematics, Hong Kong University of Science and Technology, Clear Water Bay, Kowloon, Hong Kong}
\address[HKUST2]{Department of Mechanical and Aerospace Engineering, Hong Kong University of Science and Technology, Clear Water Bay, Kowloon, Hong Kong}
\address[HKUST3]{Shenzhen Research Institute, Hong Kong University of Science and Technology, Shenzhen, China}
\address[XJTU]{Shaanxi Key Laboratory of Environment and Control for Flight Vehicle, Xi'an Jiaotong University, Xi'an, China}
\address[IAPCM]{Institute of Applied Physics and Computational Mathematics, Beijing, P.R. China}

\begin{abstract}
Conventional discrete ordinate methods are widely used for thermal radiative transfer, but their fixed number
of angular degrees of freedom (DoFs) incurs unnecessary computational cost near equilibrium.
We therefore develop a nonequilibrium-capturing unified gas-kinetic scheme (NCUGKS)
for multi-frequency radiation transport.
The scheme preserves the multiscale capability of UGKS through interface evolution
that couples transport and collision.
Radiation intensity is expanded in an orthonormal basis of the discrete angular space.
The expansion coefficients serve as angular DoFs, with their number adapted independently
to local nonequilibrium in each cell, frequency group, and quadrant.
At each time step, the coefficients to be evolved are selected using the exponentially decayed
contribution from the previous step and the requirements of interface transport.
The selected coefficients are advanced by the finite volume scheme and corrected to match
the macroscopic radiation energy before higher-degree coefficients are discarded
according to the prescribed relative tail tolerance.
Therefore, in the optically thick limit, the resulting representation retains only one angular DoF per quadrant. Numerical tests spanning different optical regimes show close agreement
with reference solutions, validating its accuracy.
Furthermore, in terms of memory and computational efficiency, the proposed method reduces memory
by factors from \(21.7\) to \(154.0\) and achieves
speedups from \(24.87\) to \(257.9\) compared with the original discrete ordinate UGKS. NCUGKS thus provides the ability to accurately simulate multiscale radiative transfer with reduced memory and computational costs.
The proposed approach also offers a framework for adapting angular degrees of freedom in other discrete ordinate methods.
\end{abstract}


\end{frontmatter}

\section{Introduction}
\label{sec:introduction}

Thermal radiative transfer is a seven dimensional problem in space, angle, frequency, and time. The discrete ordinate method (DOM) is widely used in thermal radiative transfer~\cite{LathropCarlson1966,Fiveland1984}. In conventional DOM implementations with a fixed angular quadrature, the same angular resolution is retained across transport regimes, which can increase storage and computational work in optically thick regions. Moreover, in the optically thick limit, the small photon mean free path imposes a severe time step restriction on conventional explicit DOM schemes, making the computation prohibitively expensive. These limitations call for both efficient multiscale methods and methods whose number of DoFs adapts to the local transport regime.

In the development of multiscale methods, UGKS was originally introduced for neutral gas flow~\cite{UGKS}. Its time dependent interface evolution couples transport and collision. UGKS was subsequently extended to linear radiative transfer in~\cite{Mieussens2013UGKS}. The scheme treats diffusion and free transport within a unified formulation. In the diffusion limit, the spatial mesh and time step need not resolve the photon mean free path and collision time, respectively~\cite{SunJiangXu2015}. This can reduce the computational cost in optically thick regimes. The framework was then extended to gray and multigroup thermal radiative transfer~\cite{SunJiangXu2015,SunJiangXuLi2015}. The discrete unified gas kinetic scheme (DUGKS) also couples transport and collision through a discretized integral solution along characteristics~\cite{GuoXuWang2013DUGKS}. It has been extended to radiative transfer with isotropic and anisotropic scattering~\cite{LuoWangZhangYiTan2018,SongZhangZhouGuo2020}, as well as nonlinear gray radiative transfer~\cite{SongSunGuoGao2025}. Recently, the general synthetic iterative scheme (GSIS) has been extended to multiscale gray radiative transfer and gas radiation coupling~\cite{Wang2025GSISRadiation,Zeng2026GSISGasRadiation}. Alongside these deterministic approaches, multiscale particle methods have also been developed. The asymptotic preserving Monte Carlo method combines deterministic macroscopic updates with particle transport~\cite{ShiSongXiong2023}. It converges to a diffusion discretization in the optically thick limit. The unified gas kinetic wave particle (UGKWP) method was first introduced for multiscale gas dynamics~\cite{LiuZhuXu2020UGKWP}. It describes multiscale flow through an adaptive combination of deterministic hydrodynamic waves and stochastic kinetic particles. The method has been extended to gray and frequency dependent radiative transfer, with an independent decomposition in each frequency group~\cite{LiLiuZhuZhangXu2020,YangZhuLiuXu2025}. Recent developments also include an implicit UGKWP (IUGKWP) formulation for radiative transport~\cite{LiuLiWangSongXu2023IUGKWP} and a unified gas kinetic particle formulation for radiative transfer with anisotropic scattering~\cite{HuLiuShenXiaoLi2024AnisotropicUGKP}. Numerical accuracy and the preservation of physical bounds have also motivated developments in implicit Monte Carlo (IMC) methods, including reduced teleportation error~\cite{ShiYongZhaiQiSong2018} and an iterative formulation that preserves the maximum principle for frequency dependent radiative transfer~\cite{Shi2023MPP}.

In the development of methods with adaptive DoFs, UGKWP allows the particle number to vary with the local transport regime, although statistical noise remains. Dynamical low-rank methods have also been developed to reduce the number of DoFs in radiative transfer and related linear kinetic transport equations~\cite{Brunner2027DomainDLR,Li2026EnergyStableDLR}. In this work, we develop a nonequilibrium-capturing UGKS (NCUGKS) for multi-frequency radiation transport, building on the idea introduced for neutral gas flow~\cite{LiuXu2026NCUGKSI}. Unlike the \(P_N\) moment method, which expands radiation intensity using continuous spherical harmonic functions, NCUGKS constructs its expansion directly in the discrete angular space. An orthonormal basis is defined using the discrete inner product associated with a fixed angular quadrature. The resulting basis coefficients serve as the angular DoFs and are evolved directly. This gives a new discrete basis formulation distinct from conventional DOM, which evolves nodal intensities. The angular DoFs are adapted independently for each cell, frequency group, and quadrant.

\begin{samepage}
The method is constructed from three main components.
\begin{enumerate}
 \item[(i)] \emph{Discrete basis representation and interface flux.}
 We expand the radiation intensity distribution function in a discrete orthonormal
 basis on each angular quadrant. The interface flux couples only
 coefficients associated with basis degrees that differ by at most two. This limited
 coupling enables efficient flux evaluation using the retained basis coefficients,
 without reconstructing the intensity at every angular quadrature point.

 \item[(ii)] \emph{Candidate space update.}
 The candidate space is selected to resolve both the surviving nonequilibrium
 contribution from the previous time step and interface transport. We derive a
 computable upper bound on flux truncation errors to determine the coefficients
 required by interface transport. The selected basis coefficients are then evolved
 by the finite volume scheme.

 \item[(iii)] \emph{Basis coefficient truncation.}
 After the update, higher-degree basis coefficients are discarded if their combined
 contribution does not exceed the prescribed tolerance. The retained coefficients
 are used in the next update. Our analysis shows that each quadrant retains only
 the constant basis coefficient in the optically thick limit.
\end{enumerate}
\end{samepage}

Numerical tests in different optical regimes show close agreement with reference solutions
and results from other methods. Across these tests, NCUGKS reduces memory
by factors of \(21.7\)--\(154.0\) and the mean time per
step by factors of \(24.87\)--\(257.9\) relative to the original discrete ordinate UGKS.
These results demonstrate that NCUGKS preserves the multiscale capability of UGKS
while reducing memory use and computational cost through local adaptation of the
angular DoFs.
The discrete basis representation and direct evolution of retained coefficients also offer a framework for adapting angular degrees of freedom in other discrete ordinate methods.

The remainder of the paper is organized as follows. Section~\ref{sec:model} presents the frequency dependent radiative transfer equations and their multigroup formulation. Section~\ref{sec:angular} constructs the discrete basis representation of the angular intensity, and Section~\ref{sec:ugks} derives the UGKS formulation in this basis. Section~\ref{sec:adaptation} presents the local adaptation of angular DoFs, together with an analysis of angular DoF reduction in the optically thick limit. Numerical results and conclusions are given in Sections~\ref{sec:numerics} and~\ref{sec:conclusion}, respectively.

\section{Frequency dependent thermal radiative transfer equation}
\label{sec:model}

\subsection{Governing equations}

Let \(I(t,\boldsymbol{x},\vecomega,\nu)\) be the specific intensity for
direction \(\vecomega\in\mathbb{S}^{2}\) and frequency \(\nu\). In the absence
of physical scattering, the radiation intensity and material temperature satisfy
\begin{equation}
\left\{
\begin{aligned}
 &\frac{1}{c}\frac{\partial I}{\partial t}
 +\vecomega\cdot\vgrad I
 =\sigma_{\nu}(T,\boldsymbol{x})\left(B_{\nu}(T)-I\right),\\[6pt]
 &\frac{\partial U_m(T)}{\partial t}
 =\int_{0}^{\infty}\int_{\mathbb{S}^{2}}
 \sigma_{\nu}\left(I-B_{\nu}(T)\right)\dd\vecomega\dd\nu .
\end{aligned}
\right.
\label{eq:rte}
\end{equation}
Here \(c\) is the speed of light, \(a\) the radiation constant, and
\(\sigma_{\nu}\), \(B_{\nu}\), and \(U_m(T)\) the opacity, Planck intensity,
and material internal energy density, respectively. Integrating the Planck
intensity over frequency and solid angle gives
\begin{equation}
4\pi\int_{0}^{\infty}B_{\nu}(T)\dd\nu=acT^4.
\end{equation}

\subsection{Multigroup formulation}

We discretize the finite frequency interval
\([\nu_{1/2},\nu_{G+1/2}]\) into \(G\) groups, with group \(g\)
spanning \([\nu_{g-1/2},\nu_{g+1/2}]\), \(g=1,\ldots,G\).
The group intensity is
\begin{equation}
 I_g=\int_{\nu_{g-1/2}}^{\nu_{g+1/2}}I\dd\nu.
 \label{eq:groupIntensity}
\end{equation}
The radiation density in group \(g\) is defined by
\begin{equation}
 \rho_g=\int_{\mathbb{S}^{2}}I_g\dd\vecomega.
 \label{eq:groupMoments}
\end{equation}

At equilibrium, the fraction of radiation energy in group \(g\) is
\begin{equation}
 \widetilde{\mathcal P}_g(T)
 =\frac{\displaystyle\int_{\nu_{g-1/2}}^{\nu_{g+1/2}}B_\nu(T)\dd\nu}
 {\displaystyle\int_{0}^{\infty}B_\nu(T)\dd\nu}
 =\frac{4\pi}{acT^4}
 \int_{\nu_{g-1/2}}^{\nu_{g+1/2}}B_\nu(T)\dd\nu.
 \label{eq:rawPlanckGroupWeight}
\end{equation}
Normalizing over all \(G\) groups gives
\begin{equation}
 \mathcal P_g(T)
 =\frac{\widetilde{\mathcal P}_g(T)}
 {\sum_{r=1}^{G}\widetilde{\mathcal P}_r(T)}.
 \label{eq:normalizedPlanckGroupWeight}
\end{equation}
\begin{samepage}
Thus \(\mathcal P_g(T)\geq0\) and
\(\sum_{g=1}^{G}\mathcal P_g(T)=1\). The equilibrium radiation intensity in group \(g\) is defined by
\begin{equation}
 B_g(T)=\frac{acT^4}{4\pi}\mathcal P_g(T).
 \label{eq:planckgroup}
\end{equation}
This normalization preserves
\(\sum_{g=1}^{G}4\pi B_g(T)=acT^4\) on the finite frequency interval.
\end{samepage}

For the multigroup calculations, we use logarithmically spaced frequency groups
and evaluate the Planck fractions \(\widetilde{\mathcal P}_g(T)\) by Simpson quadrature.
We construct \(G\) frequency groups based on photon energy
\(\epsilon=h_{\mathrm P}\nu\) according to
\begin{equation}
 \epsilon_{g+1/2}
 =h_{\mathrm P}\nu_{g+1/2}
 =\epsilon_{\min}
 \left(\frac{\epsilon_{\max}}{\epsilon_{\min}}\right)^{g/G},
\qquad g=0,\ldots,G.
\label{eq:logPhotonEnergyGroups}
\end{equation}
Here \(h_{\mathrm P}\) is the Planck constant,
\(\epsilon_{\min}=h_{\mathrm P}\nu_{1/2}\), and
\(\epsilon_{\max}=h_{\mathrm P}\nu_{G+1/2}\).
Temperature is expressed in energy units, with both \(\epsilon\) and \(T\)
measured in \(\mathrm{keV}\).

With
\(\epsilon_g^{\mathrm m}=(\epsilon_{g-1/2}+\epsilon_{g+1/2})/2\),
\(\Delta\epsilon_g=\epsilon_{g+1/2}-\epsilon_{g-1/2}\), and
\(\mathcal K_{\mathrm P}(\epsilon,T)
=\epsilon^3/[\exp(\epsilon/T)-1]\), Simpson quadrature in photon energy for the group integral
in Eq.~\eqref{eq:rawPlanckGroupWeight} gives
\begin{equation}
 \widetilde{\mathcal P}_g(T)
 \approx
 \frac{15\Delta\epsilon_g}{6\pi^4T^4}
 \left[
  \mathcal K_{\mathrm P}(\epsilon_{g-1/2},T)
  +4\mathcal K_{\mathrm P}(\epsilon_g^{\mathrm m},T)
  +\mathcal K_{\mathrm P}(\epsilon_{g+1/2},T)
 \right],
 \label{eq:discretePlanckGroupWeight}
\end{equation}
We approximate the group-averaged opacity using the same Simpson quadrature:
\begin{equation}
 \sigma_g(T,\boldsymbol{x})
 =\frac{1}{6}\left[
 \sigma_{\nu_{g-1/2}}(T,\boldsymbol{x})
 +4\sigma_{\nu_g^{\mathrm m}}(T,\boldsymbol{x})
 +\sigma_{\nu_{g+1/2}}(T,\boldsymbol{x})
 \right],
 \label{eq:groupOpacity}
\end{equation}
where \(\nu_g^{\mathrm m}=\epsilon_g^{\mathrm m}/h_{\mathrm P}\).
The multigroup radiation transport and material energy equations are
\begin{equation}
\left\{
\begin{aligned}
 &\frac{1}{c}\frac{\partial I_g}{\partial t}
 +\vecomega\cdot\vgrad I_g
 =\sigma_g\left(B_g-I_g\right),
 \qquad g=1,\ldots,G,\\[6pt]
 &\frac{\partial U_m(T)}{\partial t}
 =\sum_{g=1}^{G}\sigma_g
 \left(\rho_g-4\pi B_g(T)\right).
\end{aligned}
\right.
\label{eq:groupRTE}
\end{equation}

\section{Discrete basis representation of the angular intensity}
\label{sec:angular}

In this section, we expand the radiation intensity distribution function in an
orthonormal basis based on the discrete angular space. To distinguish this basis
representation from the original discrete ordinate formulation, we refer to the
latter as the nodal UGKS.

\subsection{Discrete angular space}
\label{sec:discreteAngularSpace}

On the unit sphere \(\mathbb{S}^2\), let \(\alpha\in[0,\pi]\) be the polar angle
measured from the positive \(z\) axis and \(\theta\in[0,2\pi)\) the azimuthal angle
in the \(xy\) plane. The angular direction is
\[
 \vecomega=(\Omega_x,\Omega_y,\Omega_z)^{\mathsf T}
 =(\sin\alpha\cos\theta,\sin\alpha\sin\theta,\cos\alpha)^{\mathsf T}.
\]
The solid-angle element is
\(\dd\vecomega=\sin\alpha\,\dd\alpha\,\dd\theta\).
We replace \(\alpha\) by the direction cosine
\(\mu=\cos\alpha=\Omega_z\in[-1,1]\).
In these coordinates, the angular integration measure is
\(\dd\mu\,\dd\theta\).
We then apply Gauss Legendre quadrature in \(\mu\), with nodes \(\mu_j\)
and weights \(\omega_j^\mu\), \(j=1,\ldots,N_\mu\), where \(N_\mu\) is even
and \(\mu_1<\cdots<\mu_{N_\mu}\).

\(\theta\) is discretized with \(N_\theta\) uniformly spaced midpoint nodes:
\begin{equation}
 \theta_r=\frac{2\pi(r-\tfrac12)}{N_\theta},
 \qquad r=1,\ldots,N_\theta,
 \label{eq:azimuthalQuadratureNodes}
\end{equation}
where \(N_\theta\) is a multiple of four.

For the one and two dimensional problems considered in this paper, the distribution
function is even in \(\mu\), so the contributions at \(\pm\mu_j\) are combined.
We relabel the positive direction cosines and their quadrature weights as
\(\mu_j\) and \(\omega_j^\mu\), \(j=1,\ldots,N_\mu/2\).
At each angular node \((\mu_j,\theta_r)\), the \(x\) and \(y\) components are
\begin{equation}
 (\Omega_x,\Omega_y)
 =\sqrt{1-\mu_j^2}(\cos\theta_r,\sin\theta_r),
 \label{eq:discreteAngularNodes}
\end{equation}
where \(\mu_j>0\) and \(0<\theta_r<2\pi\).

We then divide the angular space into four quadrants according to the signs of
\(\Omega_x\) and \(\Omega_y\). This partition preserves the sparse coupling of
the basis coefficients in the upwind flux at Cartesian
interfaces~\cite{LiuXu2026NCUGKSI}.
Each quadrant contains \(N_Q=N_\mu N_\theta/8\) angular nodes. The nodes in
different quadrants are related by symmetry and have identical corresponding
quadrature weights. We therefore construct the basis based on the first quadrant
angular nodes and expand the distribution independently in each quadrant using
this basis.

The positive direction cosines \(\mu_j\) and the azimuthal nodes \(0<\theta_r<\pi/2\)
form a tensor product angular grid in the first quadrant, where
\(r=1,\ldots,N_\theta/4\).
We vectorize this grid using the single index
\[
 n=(j-1)\frac{N_\theta}{4}+r.
\]
The \(x\) and \(y\) components of the \(n^{\mathrm{th}}\) angular node are denoted
by \((\eta_n,\chi_n)\).
The corresponding components in quadrant \(q\) are
\[
 \vecomega_{q,n}=(s_x^{(q)}\eta_n,s_y^{(q)}\chi_n)^{\mathsf T},
 \qquad q=0,\ldots,3,\quad n=1,\ldots,N_Q,
\]
where \(s_x^{(q)}\) and \(s_y^{(q)}\) are the signs of
\(\Omega_x\) and \(\Omega_y\), respectively, as listed below.
\begin{center}
\setlength{\tabcolsep}{16pt}
\renewcommand{\arraystretch}{1.1}
\begin{tabular}{@{}lcccc@{}}
\toprule
\(q\) & \(0\) & \(1\) & \(2\) & \(3\) \\
\midrule
\(s_x^{(q)}\) & \(+1\) & \(+1\) & \(-1\) & \(-1\) \\
\(s_y^{(q)}\) & \(+1\) & \(-1\) & \(+1\) & \(-1\) \\
\bottomrule
\end{tabular}
\end{center}

With \(\Delta\theta=2\pi/N_\theta\), symmetry in \(\mu\) gives the quadrature weights
\[
 \widetilde w_n=2\omega_j^\mu\Delta\theta
 =\frac{4\pi\omega_j^\mu}{N_\theta},
 \qquad \sum_{n=1}^{N_Q}\widetilde w_n=\pi.
\]
Normalizing by \(\pi\) gives
\begin{equation}
 w_n=\frac{\widetilde w_n}{\pi}
 =\frac{4\omega_j^\mu}{N_\theta}.
 \label{eq:referenceQuadrantWeights}
\end{equation}
Thus \(w_n>0\) and \(\sum_{n=1}^{N_Q}w_n=1\).

\(f(\vecomega_{q,n})\) denotes its common value
at the two corresponding directions on \(\mathbb{S}^2\) with opposite signs of
\(\mu\). The angular quadrature is
\begin{equation}
 \int_{\mathbb{S}^2}f\,\dd\vecomega
 \simeq
 \pi\sum_{q=0}^{3}\sum_{n=1}^{N_Q}w_n f(\vecomega_{q,n}).
 \label{eq:reducedAngularQuadrature}
\end{equation}

\subsection{Discrete basis expansion and truncation}
\label{sec:modalBasis}

For cell \(i\), group \(g\), and quadrant \(q\), let
\(\boldsymbol{I}_{i,g,q}=(I_{i,g,q,n})_{n=1}^{N_Q}\in\mathbb{R}^{N_Q}\)
denote the nodal vector of the radiation intensity distribution function.
We expand it as
\begin{equation}
 \boldsymbol{I}_{i,g,q}
 =\sum_{k=0}^{N_Q-1}a_{i,g,q,k}\boldsymbol{\psi}_k
 =\Psi\boldsymbol{a}_{i,g,q},
 \label{eq:fullModalCoordinates}
\end{equation}
where \(\boldsymbol{\psi}_k\in\mathbb{R}^{N_Q}\) are orthonormal basis vectors
with respect to the inner product defined below, and \(a_{i,g,q,k}\) are their
coefficients. We define the basis matrix by
\begin{equation}
 \Psi=[\boldsymbol{\psi}_0,\ldots,\boldsymbol{\psi}_{N_Q-1}]
 \in\mathbb{R}^{N_Q\times N_Q}.
 \label{eq:fullBasisMatrix}
\end{equation}
The corresponding coefficient vector is
\(\boldsymbol{a}_{i,g,q}=(a_{i,g,q,k})_{k=0}^{N_Q-1}\in\mathbb{R}^{N_Q}\).

For \(\boldsymbol{u},\boldsymbol{v}\in\mathbb{R}^{N_Q}\), we define the inner
product
\begin{equation}
 \langle\boldsymbol{u},\boldsymbol{v}\rangle_W
 =\boldsymbol{u}^{\mathsf T}W\boldsymbol{v},
 \qquad
 W=\operatorname{diag}(w_1,\ldots,w_{N_Q}).
 \label{eq:discreteInnerProduct}
\end{equation}
Its induced norm is
\(\|\boldsymbol{u}\|_W
=\langle\boldsymbol{u},\boldsymbol{u}\rangle_W^{1/2}\).

Orthonormality in this inner product gives
\begin{equation}
 \Psi^{\mathsf T}W\Psi=\mathbf{I}_{N_Q}.
 \label{eq:genericBasis}
\end{equation}
The coefficient projection matrix is \(\Psi^{\mathsf T}W\), giving
\begin{equation}
 \boldsymbol{a}_{i,g,q}=\Psi^{\mathsf T}W\boldsymbol{I}_{i,g,q}.
 \label{eq:fullCoefficientProjection}
\end{equation}

We construct the orthonormal basis vectors.
For \(u,v\in\mathbb{N}_0=\{0,1,2,\ldots\}\), define
\begin{equation}
 \boldsymbol{m}_{uv}
 =\bigl(\eta_1^u\chi_1^v,\ldots,\eta_{N_Q}^u\chi_{N_Q}^v\bigr)^{\mathsf T}
 \in\mathbb{R}^{N_Q}.
 \label{eq:sampledMonomialVector}
\end{equation}
The constant vector is
\[
 \boldsymbol{m}_{00}=\boldsymbol{1}=(1,\ldots,1)^{\mathsf T}.
\]
At total degree one, the vectors are
\[
 \boldsymbol{m}_{10}=(\eta_1,\ldots,\eta_{N_Q})^{\mathsf T},\qquad
 \boldsymbol{m}_{01}=(\chi_1,\ldots,\chi_{N_Q})^{\mathsf T}.
\]
At total degree two, they are
\[
 \boldsymbol{m}_{20}=(\eta_1^2,\ldots,\eta_{N_Q}^2)^{\mathsf T},\quad
 \boldsymbol{m}_{11}=(\eta_1\chi_1,\ldots,\eta_{N_Q}\chi_{N_Q})^{\mathsf T},\quad
 \boldsymbol{m}_{02}=(\chi_1^2,\ldots,\chi_{N_Q}^2)^{\mathsf T}.
\]
These vectors are generally not orthogonal. We apply Gram Schmidt in
increasing total degree \(u+v\), with \(u\) decreasing within each degree.
Since \(\sum_{n=1}^{N_Q}w_n=1\), the constant vector gives the first basis vector:
\[
 \boldsymbol{\psi}_0
 =\frac{\boldsymbol{m}_{00}}{\|\boldsymbol{m}_{00}\|_W}
 =\boldsymbol{1}.
\]
Let \(k\) be the number of basis vectors already constructed. For the next
candidate \(\boldsymbol{m}_{uv}\), we remove its projections onto
\(\boldsymbol{\psi}_0,\ldots,\boldsymbol{\psi}_{k-1}\):
\begin{equation}
 \boldsymbol{m}_{uv}^{\perp}
 =\boldsymbol{m}_{uv}
 -\sum_{j=0}^{k-1}
 \langle\boldsymbol{\psi}_j,\boldsymbol{m}_{uv}\rangle_W
 \boldsymbol{\psi}_j.
 \label{eq:weightedGramSchmidtResidual}
\end{equation}
If \(\boldsymbol{m}_{uv}\) is a linear combination of the existing basis vectors,
then \(\boldsymbol{m}_{uv}^{\perp}=\boldsymbol{0}\) and \(\boldsymbol{m}_{uv}\)
is skipped. Otherwise, normalization gives the next basis vector:
\begin{equation}
 \boldsymbol{\psi}_k
 =\frac{\boldsymbol{m}_{uv}^{\perp}}
 {\|\boldsymbol{m}_{uv}^{\perp}\|_W}.
 \label{eq:weightedGramSchmidtNormalization}
\end{equation}
We assign the discrete basis vector \(\boldsymbol\psi_k\) the degree label
\[
 \deg_\Psi(k):=u+v,
\]
inherited from its generating monomial \(\eta^u\chi^v\).
The coefficient \(a_{i,g,q,k}\) carries the same degree label.

The construction stops once \(N_Q\) basis vectors have been obtained.
The truncation degree \(p\) specifies that all basis vectors with degree at most
\(p\) are retained. Their number, also the number of angular DoFs in one quadrant, is
\[
 D_p=1+\max\{k:\deg_\Psi(k)\leq p\},
 \qquad p\in\mathbb N_0,
\]
where \(0\leq k<N_Q\).
The highest basis degree is \mbox{\(p_{\mathrm{full}}=\deg_\Psi(N_Q-1)\)}, with
\(D_{p_{\mathrm{full}}}=N_Q\).
Algorithm~\ref{alg:modalBasisConstruction} summarizes the construction.
\begin{algorithm}[H]
\caption{Construction of the discrete basis}
\label{alg:modalBasisConstruction}
\begin{algorithmic}[1]
\State \textbf{Input}\quad Quadrature data
\(\{(\eta_n,\chi_n,w_n)\}_{n=1}^{N_Q}\)
\State \textbf{Output}\quad Full basis \(\Psi\), basis degrees
\(\{\deg_\Psi(k)\}_{k=0}^{N_Q-1}\), and
cumulative dimensions \(\{D_r\}\)
\State \(W\gets\operatorname{diag}(w_1,\ldots,w_{N_Q})\)
\State \(k\gets0\), \(r\gets0\)
\While{\(k<N_Q\)}
 \For{\(u=r,r-1,\ldots,0\)}
  \State \(v\gets r-u\)
  \State
  \(\boldsymbol{m}_{uv}\gets
  (\eta_n^u\chi_n^v)_{n=1}^{N_Q}\)
  \State
  \(\displaystyle
  \boldsymbol{m}_{uv}^{\perp}\gets
  \boldsymbol{m}_{uv}
  -\sum_{j=0}^{k-1}
  \langle\boldsymbol{\psi}_j,\boldsymbol{m}_{uv}\rangle_W
  \boldsymbol{\psi}_j\)
  \If{\(\boldsymbol{m}_{uv}^{\perp}\neq\boldsymbol{0}\)}
   \State
   \(\displaystyle
   \boldsymbol{\psi}_k\gets
   \boldsymbol{m}_{uv}^{\perp}/
   \|\boldsymbol{m}_{uv}^{\perp}\|_W\)
   \State \(\deg_\Psi(k)\gets r\), \(k\gets k+1\)
  \EndIf
 \EndFor
 \State \(D_r\gets k\), \(r\gets r+1\)
\EndWhile
\State \(\Psi\gets
[\boldsymbol{\psi}_0,\ldots,\boldsymbol{\psi}_{N_Q-1}]\)
\end{algorithmic}
\end{algorithm}

With truncation degree \(p\), the distribution is represented by
\begin{equation}
 \boldsymbol I_{i,g,q}^{(p)}
 =\sum_{k=0}^{D_p-1}a_{i,g,q,k}\boldsymbol\psi_k.
 \label{eq:truncatedModalMaps}
\end{equation}
The retained coefficients are
\begin{equation}
 a_{i,g,q,k}
 =\langle\boldsymbol\psi_k,\boldsymbol I_{i,g,q}\rangle_W,
 \qquad 0\leq k<D_p,
 \label{eq:truncatedModalCoefficients}
\end{equation}
forming the vector
\(\boldsymbol a_{i,g,q}^{(p)}
=(a_{i,g,q,0},\ldots,a_{i,g,q,D_p-1})^{\mathsf T}
\in\mathbb R^{D_p}\).
The discarded coefficients give the truncation error
\[
 \boldsymbol{I}_{i,g,q}-\boldsymbol{I}_{i,g,q}^{(p)}
 =\sum_{k=D_p}^{N_Q-1}a_{i,g,q,k}\boldsymbol{\psi}_k.
\]
By orthonormality,
\begin{equation}
 \left\|\boldsymbol{I}_{i,g,q}
  -\boldsymbol{I}_{i,g,q}^{(p)}\right\|_W^2
 =\sum_{k=D_p}^{N_Q-1}a_{i,g,q,k}^2.
 \label{eq:modalParseval}
\end{equation}
\begin{remark}
Equation~\eqref{eq:modalParseval} enables efficient evaluation of the truncation
error directly from the discarded coefficients, without reconstructing the nodal
distribution.
\end{remark}

\subsection{Discrete angular moments and multiplication matrices}
\label{sec:modalOperators}

We first evaluate the radiation density from the basis coefficients.
The cell index \(i\) is suppressed in this subsection.

\begin{proposition}[Discrete radiation density]
\label{prop:modalRadiationDensity}
For frequency group \(g\), let \(\boldsymbol I_{g,q}^{(p_q)}\) denote the
truncated basis representation in quadrant \(q\), with \(p_q\geq0\).
The discrete radiation density satisfies
\begin{equation}
 \rho_g=\pi\sum_{q=0}^{3}a_{g,q,0}.
 \label{eq:modalGroupEnergy}
\end{equation}
Thus basis truncation preserves the discrete radiation density.
\end{proposition}

\begin{proof}
Since \(\boldsymbol\psi_0=\boldsymbol1\), orthonormality gives
\[
 \sum_{n=1}^{N_Q}w_n(\boldsymbol\psi_k)_n
 =\langle\boldsymbol\psi_0,\boldsymbol\psi_k\rangle_W
 =\begin{cases}
   1, & k=0,\\
   0, & k\geq1.
  \end{cases}
\]
Substituting the basis expansion into the angular quadrature and interchanging
 the finite sums yields
\[
 \begin{aligned}
  \rho_g
  &=\pi\sum_{q=0}^{3}\sum_{n=1}^{N_Q}w_n I_{g,q,n}^{(p_q)}\\
  &=\pi\sum_{q=0}^{3}\sum_{n=1}^{N_Q}w_n
    \sum_{k=0}^{D_{p_q}-1}a_{g,q,k}(\boldsymbol\psi_k)_n\\
  &=\pi\sum_{q=0}^{3}\sum_{k=0}^{D_{p_q}-1}a_{g,q,k}
    \sum_{n=1}^{N_Q}w_n(\boldsymbol\psi_k)_n\\
  &=\pi\sum_{q=0}^{3}a_{g,q,0}.
 \end{aligned}
\]
The constant coefficient is retained for every \(p_q\geq0\), so truncation
does not change this sum.
\end{proof}

\begin{remark}
Given the basis coefficients, evaluating \(\rho_g\) requires only the four
constant coefficients. The cost per cell and frequency group is therefore
\(O(1)\), compared with \(O(N_\mu N_\theta)\) for nodal quadrature.
\end{remark}

For a nodal distribution vector \(\boldsymbol I=(I_n)_{n=1}^{N_Q}\),
multiplication by \(\eta\) has the vector form
\[
 (\eta_1 I_1,\ldots,\eta_{N_Q}I_{N_Q})^{\mathsf T}=E\boldsymbol I,
\]
where \(E=\operatorname{diag}(\eta_1,\ldots,\eta_{N_Q})\).
For \(\boldsymbol I=\Psi\boldsymbol a\), define the linear multiplication
matrix by
\begin{equation}
 J^\eta=\Psi^{\mathsf T}WE\Psi.
 \label{eq:linearModalOperators}
\end{equation}
Its entries are
\[
 (J^\eta)_{\ell k}
 =\langle\boldsymbol\psi_\ell,E\boldsymbol\psi_k\rangle_W,
 \qquad 0\leq\ell,k<N_Q.
\]
Then \(\Psi^{\mathsf T}W(E\boldsymbol I)=J^\eta\boldsymbol a\).
The matrices \(C\) and \(J^\chi\) for multiplication by \(\chi\) are defined analogously.
The corresponding quadratic matrices are
\begin{equation}
 K^{\eta\eta}=\Psi^{\mathsf T}WE^2\Psi,
 \qquad
 K^{\eta\chi}=\Psi^{\mathsf T}WEC\Psi,
 \qquad
 K^{\chi\chi}=\Psi^{\mathsf T}WC^2\Psi.
 \label{eq:modalOperators}
\end{equation}
The following proposition characterizes the degree coupling of linear angular multiplication.

\begin{proposition}[Block structure of the linear multiplication matrices]
\label{prop:modalBlockStructure}
In the degree ordering, the linear multiplication matrices satisfy
\begin{equation}
 J^\eta_{st}=J^\chi_{st}=0\qquad\text{for }|s-t|>1.
 \label{eq:coordinateBlockBandwidth}
\end{equation}
Here the \((s,t)\) block maps coefficients of degree \(t\) to those of degree
\(s\), with \(0\leq s,t\leq p_{\mathrm{full}}\).
\end{proposition}

\begin{proof}
Let \(\boldsymbol\psi_k\) and \(\boldsymbol\psi_\ell\) have degrees
\(t\) and \(s\), respectively.
By the construction in Section~\ref{sec:modalBasis},
\(\boldsymbol\psi_k\) contains the nodal values of a polynomial
of total degree at most \(t\). Since
\((E\boldsymbol\psi_k)_n=\eta_n(\boldsymbol\psi_k)_n\)
and \(\eta\) is linear, it follows that
\[
 E\boldsymbol\psi_k\in
 \operatorname{span}\{\boldsymbol\psi_0,\ldots,
 \boldsymbol\psi_{D_{t+1}-1}\},
\]
where the indicated basis vectors span the nodal values of all
polynomials of total degree at most \(t+1\).
Orthogonality therefore gives
\[
 \langle\boldsymbol\psi_\ell,E\boldsymbol\psi_k\rangle_W=0,
 \qquad s>t+1.
\]
Since \(E^{\mathsf T}W=WE\), the same argument applied to
\(E\boldsymbol\psi_\ell\) yields
\[
 \langle\boldsymbol\psi_\ell,E\boldsymbol\psi_k\rangle_W
 =\langle E\boldsymbol\psi_\ell,\boldsymbol\psi_k\rangle_W=0,
 \qquad t>s+1.
\]
Thus \(J^\eta_{st}=0\) for \(|s-t|>1\).
The same argument applies to \(J^\chi\).
\end{proof}

Consequently, \(J^\eta\) and \(J^\chi\) are block tridiagonal.
Analogously,
\[
 K^{\eta\eta}_{st}=K^{\eta\chi}_{st}=K^{\chi\chi}_{st}=0,
 \qquad |s-t|>2,
\]
so \(K^{\eta\eta}\), \(K^{\eta\chi}\), and \(K^{\chi\chi}\) are block
pentadiagonal.
For a fixed quadrature, the discrete basis and multiplication matrices are
precomputed.

\section{UGKS formulation in the discrete basis}
\label{sec:ugks}

\subsection{Original UGKS}
\label{sec:originalUGKS}

We first recall the multigroup UGKS for thermal radiative transfer developed by
Sun et al.~\cite{SunJiangXuLi2015}.
At each Cartesian face \(f\), the unit normal \(\boldsymbol n_f\) points from
cell \(L\) to cell \(R\) along the positive coordinate axis.
Omitting \(g\), linear reconstruction from the cell centers
\(\boldsymbol x_L,\boldsymbol x_R\) to the face center \(\boldsymbol x_f\) gives
\[
 B_f^L=B_L+\nabla_hB_L\cdot(\boldsymbol{x}_f-\boldsymbol{x}_L),
 \qquad
 B_f^R=B_R+\nabla_hB_R\cdot(\boldsymbol{x}_f-\boldsymbol{x}_R).
\]
For the opacity,
\[
 \sigma_f^L=\sigma_L+\nabla_h\sigma_L\cdot
 (\boldsymbol{x}_f-\boldsymbol{x}_L),
 \qquad
 \sigma_f^R=\sigma_R+\nabla_h\sigma_R\cdot
 (\boldsymbol{x}_f-\boldsymbol{x}_R).
\]
The interface values are
\begin{equation}
 B_f=\frac{B_f^L+B_f^R}{2},
 \qquad
 \sigma_f=\frac{2\sigma_f^L\sigma_f^R}{\sigma_f^L+\sigma_f^R}.
 \label{eq:faceEquilibriumOpacity}
\end{equation}

At a vertical face,
\begin{equation}
 (\partial_xB)_f=\frac{B_R-B_L}{\Delta x},
 \qquad
 (\partial_yB)_f
 =\frac{(\partial_y^hB)_L+(\partial_y^hB)_R}{2}.
 \label{eq:faceEquilibriumGradient}
\end{equation}
The horizontal-face formula follows by interchanging \(x\) and \(y\).

We index the discrete angular directions by \(m\), or by \((q,\kappa)\)
in quadrant notation, where \(q=0,\ldots,3\) and \(\kappa=1,\ldots,N_Q\).
Here \(\kappa\) replaces the angular index \(n\) used in
Section~\ref{sec:discreteAngularSpace} to distinguish it from the time level \(n\).
The normal component of ordinate \(m\) is
\[
 \Omega_{f,m}=\vecomega_m\cdot\boldsymbol{n}_f.
\]
The relaxation time for group \(g\) is
\begin{equation}
 \tau_{g,f}=\frac{1}{c\sigma_{g,f}}.
 \label{eq:faceParameters}
\end{equation}
With \(\tau=\tau_{g,f}\), the characteristic solution of the radiation transport
equation in Eq.~\eqref{eq:groupRTE} at face \(f\) is
\begin{align}
 I_g(\boldsymbol{x}_f,t^n+t,\vecomega_m)
 ={}&\frac{1}{\tau}\int_0^t e^{-(t-s)/\tau}
 B_g\bigl(\boldsymbol{x}_f-c\vecomega_m(t-s),t^n+s\bigr)\,\dd s
 \nonumber\\
 &+e^{-t/\tau}
 I_g^n(\boldsymbol{x}_f-c\vecomega_m t,\vecomega_m),
 \label{eq:integralsolution}
\end{align}
The first-order
space time expansion along the characteristic is
\begin{equation}
 B_g\bigl(\boldsymbol{x}_f-c\vecomega_m(t-s),t^n+s\bigr)
 \approx
 B_{g,f}-c(t-s)\vecomega_m\cdot\nabla B_{g,f}
 +s\partial_tB_{g,f}.
 \label{eq:equilibriumSpaceTimeExpansion}
\end{equation}

The left and right linear reconstructions at face \(\Gamma_f\) are
\begin{equation}
 \begin{aligned}
  I_{g,f,m}^{L,n}
  &=I_{L,g,m}^{n}
    +\nabla I_{L,g,m}^{n}\cdot
     (\boldsymbol{x}_f-\boldsymbol{x}_L),\\
  I_{g,f,m}^{R,n}
  &=I_{R,g,m}^{n}
    +\nabla I_{R,g,m}^{n}\cdot
     (\boldsymbol{x}_f-\boldsymbol{x}_R).
 \end{aligned}
 \label{eq:leftRightFaceReconstruction}
\end{equation}
The corresponding gradients are
\[
 \nabla I_{g,f,m}^{L,n}=\nabla I_{L,g,m}^{n},
 \qquad
 \nabla I_{g,f,m}^{R,n}=\nabla I_{R,g,m}^{n}.
\]
Let \(I_{g,f,m}^{\mathrm{up},n}\) and
\(\nabla I_{g,f,m}^{\mathrm{up},n}\) denote the \(L\) reconstruction for
\(\Omega_{f,m}>0\) and the \(R\) reconstruction otherwise. The initial value
in Eq.~\eqref{eq:integralsolution} is
\begin{equation}
 I_g^n(\boldsymbol{x}_f-c\vecomega_m t,\vecomega_m)
 =I_{g,f,m}^{\mathrm{up},n}
 -ct\vecomega_m\cdot\nabla I_{g,f,m}^{\mathrm{up},n}.
 \label{eq:leftRightCharacteristicValue}
\end{equation}

Substitution of Eqs.~\eqref{eq:equilibriumSpaceTimeExpansion} and
\eqref{eq:leftRightCharacteristicValue} into Eq.~\eqref{eq:integralsolution}
gives
\begin{align}
 I_g(\boldsymbol{x}_f,t^n+t,\vecomega_m)
 ={}&\gamma_1B_{g,f}
 +\gamma_2\vecomega_m\cdot\nabla B_{g,f}
 +\gamma_3\partial_tB_{g,f}
 \nonumber\\
 &+\gamma_4 I_{g,f,m}^{\mathrm{up},n}
 +\gamma_5\vecomega_m\cdot\nabla I_{g,f,m}^{\mathrm{up},n},
 \label{eq:nodalInterfaceIntensity}
\end{align}
The coefficients in Eq.~\eqref{eq:nodalInterfaceIntensity} are
\begin{equation}
 \begin{aligned}
  \gamma_1&=1-e^{-t/\tau},
  &\gamma_2&=-c\tau\left[1-\left(1+\frac{t}{\tau}\right)e^{-t/\tau}\right],\\
  \gamma_3&=t-\tau+\tau e^{-t/\tau},
  &\gamma_4&=e^{-t/\tau},\\
  \gamma_5&=-ct e^{-t/\tau}.
 \end{aligned}
 \label{eq:interfaceTimeCoefficients}
\end{equation}
Set \(\partial_tB_{g,f}=0\). The
time-averaged microscopic flux at ordinate \(m\) is defined by
\begin{equation}
 \mathcal{F}_{g,f,m}
 =\frac{1}{\Delta t}\int_0^{\Delta t}
 c\Omega_{f,m}I_g(\boldsymbol{x}_f,t^n+t,\vecomega_m)\,\dd t.
 \label{eq:nodalMicroscopicFluxDefinition}
\end{equation}
Define the dimensionless face optical thickness \(\Lambda\) and the auxiliary
function \(\phi\) by
\begin{equation}
 \Lambda=\frac{\Delta t}{\tau},
 \qquad
 \phi=\frac{1-e^{-\Lambda}}{\Lambda}.
\label{eq:interfaceScaleFunction}
\end{equation}
Substituting Eq.~\eqref{eq:nodalInterfaceIntensity} into
Eq.~\eqref{eq:nodalMicroscopicFluxDefinition} and integrating in time gives
the coefficients
\begin{align}
 \mathcal C_1
 &=\frac{c}{\Delta t}\int_0^{\Delta t}\gamma_1\,\dd t
 =c(1-\phi),
 &
 \mathcal C_2
 &=\frac{c}{\Delta t}\int_0^{\Delta t}\gamma_2\,\dd t
 =-c^2\Delta t\frac{1+e^{-\Lambda}-2\phi}{\Lambda},
 \label{eq:q12}\\
 \mathcal C_3
 &=\frac{c}{\Delta t}\int_0^{\Delta t}\gamma_4\,\dd t
 =c\phi,
 &
 \mathcal C_4
 &=\frac{c}{\Delta t}\int_0^{\Delta t}\gamma_5\,\dd t
 =-c^2\Delta t
 \frac{1-e^{-\Lambda}-\Lambda e^{-\Lambda}}{\Lambda^2}.
 \label{eq:q34}
\end{align}
The microscopic flux is therefore
\begin{align}
 \mathcal{F}_{g,f,m}
 ={}&\mathcal C_1\Omega_{f,m}B_{g,f}
 +\mathcal C_2\Omega_{f,m}\vecomega_m\cdot\nabla B_{g,f}
 \nonumber\\
 &+\mathcal C_3\Omega_{f,m}I_{g,f,m}^{\mathrm{up},n}
 +\mathcal C_4\Omega_{f,m}\vecomega_m\cdot
 \nabla I_{g,f,m}^{\mathrm{up},n},
 \label{eq:UGKSflux}
\end{align}
At face \(f\), set
\(\Omega_{f,q,\kappa}=\vecomega_{q,\kappa}\cdot\boldsymbol{n}_f\) and define
\begin{equation}
 \begin{aligned}
  \mathcal G_{3,g,f}^{n}
  &=\pi\sum_{q=0}^{3}\sum_{\kappa=1}^{N_Q}
    w_\kappa\Omega_{f,q,\kappa}I_{g,f,q,\kappa}^{\mathrm{up},n},\\
  \mathcal G_{4,g,f}^{n}
  &=\pi\sum_{q=0}^{3}\sum_{\kappa=1}^{N_Q}
    w_\kappa\Omega_{f,q,\kappa}\vecomega_{q,\kappa}\cdot
    \nabla I_{g,f,q,\kappa}^{\mathrm{up},n}.
 \end{aligned}
 \label{eq:frozenAngularMoments}
\end{equation}
Taking the discrete angular moment of Eq.~\eqref{eq:UGKSflux} gives the
macroscopic group energy flux
\begin{equation}
 \mathcal{F}_{g,f}^{\rho}
 =\frac{4\pi}{3}\mathcal C_2\boldsymbol{n}_f\cdot\nabla B_{g,f}
 +\mathcal C_3\mathcal G_{3,g,f}^{n}
 +\mathcal C_4\mathcal G_{4,g,f}^{n}.
 \label{eq:nodalMacroscopicFluxExpansion}
\end{equation}

\subsection{Coupled update of radiation and material temperature}
\label{sec:coupledUGKSUpdate}

For cell \(V_i\), define
\begin{equation}
 \begin{aligned}
  \sigma_{i,g}^{n+1}
  &=\sigma_{i,g}(T_i^{n+1}),
  &\mathcal P_{i,g}^{n+1}
  &=\mathcal P_g(T_i^{n+1}),
  &B_{i,g}^{n+1}
  &=\frac{ac(T_i^{n+1})^4}{4\pi}\mathcal P_{i,g}^{n+1},\\
  \vartheta_{i,g}^{n+1}
  &=\Delta t\,\sigma_{i,g}^{n+1},
  &\lambda_{i,g}^{n+1}
  &=c\vartheta_{i,g}^{n+1}.
 \end{aligned}
 \label{eq:collisionParameters}
\end{equation}
The superscript \(n\) denotes the same quantities evaluated at \(T_i^n\).
Define \(s_{i,f}\in\{-1,1\}\) so that \(s_{i,f}\boldsymbol n_f\) is the
outward normal to \(V_i\). At an interior face, \(s_{L,f}=1\) and
\(s_{R,f}=-1\). The microscopic intensity update is
\begin{equation}
 \begin{aligned}
 \left(1+\lambda_{i,g}^{n+1}\right)I_{i,g,m}^{n+1}
 ={}&I_{i,g,m}^{n}
 -\frac{\Delta t}{|V_i|}
 \sum_{\Gamma_f\subset\partial V_i}
 s_{i,f}|\Gamma_f|\,\mathcal{F}_{g,f,m}
 +\lambda_{i,g}^{n+1}B_{i,g}^{n+1},
 \end{aligned}
 \label{eq:implicitNodalUpdate}
\end{equation}

Taking its discrete angular moment gives the group radiation energy update
\begin{equation}
 \begin{aligned}
 \left(1+\lambda_{i,g}^{n+1}\right)\rho_{i,g}^{n+1}
 ={}&\rho_{i,g}^{n}
 -\frac{\Delta t}{|V_i|}
  \sum_{\Gamma_f\subset\partial V_i}
  s_{i,f}|\Gamma_f|\,\mathcal{F}_{g,f}^{\rho}\\
 &+4\pi\lambda_{i,g}^{n+1}B_{i,g}^{n+1}.
 \end{aligned}
 \label{eq:implicitGroupEnergyUpdate}
\end{equation}
This equation is coupled to the material energy balance
\begin{equation}
 U_m(T_i^{n+1})-U_m(T_i^n)
 =\sum_{g=1}^{G}\vartheta_{i,g}^{n+1}
 \left(\rho_{i,g}^{n+1}-4\pi B_{i,g}^{n+1}\right).
 \label{eq:implicitMaterialUpdate}
\end{equation}
Substituting Eq.~\eqref{eq:implicitGroupEnergyUpdate} into
Eq.~\eqref{eq:implicitMaterialUpdate} eliminates \(\rho_{i,g}^{n+1}\) and
gives
\begin{equation}
 U_m(T_i^{n+1})
 +ac\mathcal D_i^{n+1}(T_i^{n+1})^4
 =
 U_m(T_i^n)+A_i^{n+1},
 \label{eq:coupledTemperatureEquation}
\end{equation}
where
\begin{equation}
 \begin{aligned}
  A_i^{n+1}
  &=
  \sum_{g=1}^{G}
  \frac{\vartheta_{i,g}^{n+1}}{1+c\vartheta_{i,g}^{n+1}}
  \left[
   \rho_{i,g}^{n}
   -\frac{\Delta t}{|V_i|}
    \sum_{\Gamma_f\subset\partial V_i}
    s_{i,f}|\Gamma_f|\,\mathcal{F}_{g,f}^{\rho}
  \right],\\
  \mathcal D_i^{n+1}
  &=
  \sum_{g=1}^{G}
  \frac{\vartheta_{i,g}^{n+1}\mathcal P_{i,g}^{n+1}}
  {1+c\vartheta_{i,g}^{n+1}}.
 \end{aligned}
 \label{eq:coupledTemperatureCoefficients}
\end{equation}
We solve the coupled radiation and material energy equations~\eqref{eq:implicitGroupEnergyUpdate}
and~\eqref{eq:implicitMaterialUpdate} by Picard iteration. At iteration \(\ell\), the coefficients and macroscopic flux
are evaluated from the current temperatures \(\{T_i^{(\ell)}\}\), while the angular
moments in Eq.~\eqref{eq:frozenAngularMoments} remain fixed.
Algorithm~\ref{alg:originalUGKSTimeStep} summarizes one time step of the
original multigroup UGKS.

\begin{algorithm}[H]
\caption{One time step of the original multigroup UGKS}
\label{alg:originalUGKSTimeStep}
\begin{algorithmic}[1]
\State \textbf{Given}\quad
\(\{I_{i,g,m}^{n}\}\), \(\{\rho_{i,g}^{n}\}\), \(\{T_i^n\}\),
\(\Delta t\), and iteration tolerances \(\varepsilon_T,\varepsilon_\rho>0\)
\State Reconstruct \(I^n\) at the faces and form
\(\mathcal G_{3,g,f}^{n}\) and \(\mathcal G_{4,g,f}^{n}\) from
Eq.~\eqref{eq:frozenAngularMoments}
\State \(T_i^{(0)}\gets T_i^n\),
\(\rho_{i,g}^{(0)}\gets\rho_{i,g}^n\), and \(\ell\gets0\)
\Repeat
 \State Evaluate
 \(\sigma_{i,g}^{(\ell)}=\sigma_{i,g}(T_i^{(\ell)})\),
 \(\mathcal P_{i,g}^{(\ell)}=\mathcal P_g(T_i^{(\ell)})\), and
 \(B_{i,g}^{(\ell)}=ac(T_i^{(\ell)})^4
 \mathcal P_{i,g}^{(\ell)}/(4\pi)\)
 \State Form \(\vartheta_{i,g}^{(\ell)}\) and
 \(\lambda_{i,g}^{(\ell)}\) from Eq.~\eqref{eq:collisionParameters}
 \State Form \(B_{g,f}^{(\ell)}\), \(\nabla B_{g,f}^{(\ell)}\), and
 \(\sigma_{g,f}^{(\ell)}\) using
 Eqs.~\eqref{eq:faceEquilibriumOpacity} and
 \eqref{eq:faceEquilibriumGradient}
 \State Evaluate \(\mathcal C_{j,f,g}^{(\ell)}\), \(j=1,\ldots,4\), from
 Eqs.~\eqref{eq:q12}--\eqref{eq:q34}
 \State Form \(\mathcal F_{g,f}^{\rho,(\ell)}\) from
 Eq.~\eqref{eq:nodalMacroscopicFluxExpansion} using the fixed
 \(\mathcal G_{3,g,f}^{n}\) and \(\mathcal G_{4,g,f}^{n}\)
 \State Form \(A_i^{(\ell)}\) and \(\mathcal D_i^{(\ell)}\) from
 Eq.~\eqref{eq:coupledTemperatureCoefficients}
 \ForAll{cells \(i\)}
  \State Solve
  \(U_m(T)+ac\mathcal D_i^{(\ell)}T^4
  =U_m(T_i^n)+A_i^{(\ell)}\)
  for \(T_i^{(\ell+1)}\)
  \State Recover \(\rho_{i,g}^{(\ell+1)}\) from
  Eq.~\eqref{eq:implicitGroupEnergyUpdate} using \(T_i^{(\ell+1)}\) and the
  coefficients evaluated at iteration \(\ell\)
 \EndFor
 \State \(\ell\gets\ell+1\)
\Until{\(\max_i|T_i^{(\ell)}-T_i^{(\ell-1)}|\leq\varepsilon_T\)
 and
 \(\max_{i,g}|\rho_{i,g}^{(\ell)}-\rho_{i,g}^{(\ell-1)}|
 \leq\varepsilon_\rho\)}
\State Set \(T_i^{n+1}\gets T_i^{(\ell)}\) and
\(\rho_{i,g}^{n+1}\gets\rho_{i,g}^{(\ell)}\); retain the macroscopic flux
\(\mathcal F_{g,f}^{\rho,(\ell-1)}\) from the final Picard sweep
\State Update the cell- and face-centered equilibrium quantities and
\(\mathcal C_{j,f,g}\) from \(T_i^{n+1}\)
\State Evaluate \(\mathcal F_{g,f,m}\) once from Eq.~\eqref{eq:UGKSflux};
recover \(I_{i,g,m}^{n+1}\) from Eq.~\eqref{eq:implicitNodalUpdate}, imposing
\(\rho_{i,g}^{n+1}\) as its discrete angular moment
\end{algorithmic}
\end{algorithm}

\subsection{Basis representation of the UGKS}
\label{sec:modalUGKS}

For fixed \(g\), \(f\), and \(q\), set
\[
 \begin{aligned}
\boldsymbol{I}(t)
 &=\Bigl(I_g(\boldsymbol{x}_f,t^n+t,\vecomega_{q,\kappa})\Bigr)_{\kappa=1}^{N_Q},\\
\boldsymbol{B}(t,s)
 &=\Bigl(B_g(\boldsymbol{x}_f-c\vecomega_{q,\kappa}(t-s),t^n+s)\Bigr)_{\kappa=1}^{N_Q},\\
\boldsymbol{I}^{n}(t)
 &=\Bigl(I_g^n(\boldsymbol{x}_f-c\vecomega_{q,\kappa}t,
                   \vecomega_{q,\kappa})\Bigr)_{\kappa=1}^{N_Q}.
 \end{aligned}
\]
With \(\boldsymbol{a}(t)=\Psi^{\mathsf T}W\boldsymbol{I}(t)\),
Eq.~\eqref{eq:integralsolution} gives
\begin{equation}
 \begin{aligned}
 \boldsymbol{a}(t)
 &=\frac{1}{\tau}\int_0^t e^{-(t-s)/\tau}
 \Psi^{\mathsf T}W\boldsymbol{B}(t,s)\,\dd s
 +e^{-t/\tau}\Psi^{\mathsf T}W
 \boldsymbol{I}^{n}(t).
 \end{aligned}
 \label{eq:modalIntegralSolution}
\end{equation}

At a Cartesian face, all ordinates in quadrant \(q\) use the same upwind
side. Using the reconstructed values from Section~\ref{sec:originalUGKS},
define
\begin{equation}
 \boldsymbol a^{\mathrm{up},n}
 =\Psi^{\mathsf T}W\boldsymbol I^{\mathrm{up},n}.
 \label{eq:modalUpwindCoefficients}
\end{equation}
The corresponding spatial slopes are
\begin{equation}
 \begin{aligned}
  \boldsymbol a_x^{\mathrm{up},n}
  &=\Psi^{\mathsf T}W\partial_x\boldsymbol I^{\mathrm{up},n},\\[3pt]
  \boldsymbol a_y^{\mathrm{up},n}
  &=\Psi^{\mathsf T}W\partial_y\boldsymbol I^{\mathrm{up},n}.
 \end{aligned}
 \label{eq:modalUpwindSlopes}
\end{equation}
Let \(\boldsymbol e_0=(1,0,\ldots,0)^{\mathsf T}\in\mathbb R^{N_Q}\).
Using \(\Psi^{\mathsf T}W\boldsymbol{1}=\boldsymbol{e}_0\) and the angular
multiplication matrices in Eqs.~\eqref{eq:linearModalOperators} and~\eqref{eq:modalOperators}, the two projected terms in
Eq.~\eqref{eq:modalIntegralSolution} are
\begin{align}
 \Psi^{\mathsf T}W\boldsymbol{B}(t,s)
 \approx{}&
 B_{g,f}\boldsymbol{e}_0
 -c(t-s)\left[
 s_x^{(q)}(\partial_xB_{g,f})J^\eta\boldsymbol{e}_0
 +s_y^{(q)}(\partial_yB_{g,f})J^\chi\boldsymbol{e}_0
 \right]
 +s(\partial_tB_{g,f})\boldsymbol{e}_0,
 \label{eq:projectedEquilibriumExpansion}\\
 \Psi^{\mathsf T}W\boldsymbol{I}^{n}(t)
 ={}&
 \boldsymbol{a}^{\mathrm{up},n}
 -ct\left[
 s_x^{(q)}J^\eta\boldsymbol{a}_{x}^{\mathrm{up},n}
 +s_y^{(q)}J^\chi\boldsymbol{a}_{y}^{\mathrm{up},n}
 \right].
 \label{eq:projectedCharacteristicValue}
\end{align}
Substitution of Eqs.~\eqref{eq:projectedEquilibriumExpansion} and
\eqref{eq:projectedCharacteristicValue} into
Eq.~\eqref{eq:modalIntegralSolution} yields
\begin{align}
 \boldsymbol{a}(t)
 ={}&
 \gamma_1B_{g,f}\boldsymbol{e}_0
 +\gamma_2\left[
 s_x^{(q)}(\partial_xB_{g,f})J^\eta\boldsymbol{e}_0
 +s_y^{(q)}(\partial_yB_{g,f})J^\chi\boldsymbol{e}_0
 \right]
 +\gamma_3(\partial_tB_{g,f})\boldsymbol{e}_0
 \nonumber\\
 &+\gamma_4\boldsymbol{a}^{\mathrm{up},n}
 +\gamma_5\left[
 s_x^{(q)}J^\eta\boldsymbol{a}_{x}^{\mathrm{up},n}
 +s_y^{(q)}J^\chi\boldsymbol{a}_{y}^{\mathrm{up},n}
 \right].
 \label{eq:modalInterfaceIntensity}
\end{align}

For \(d\in\{x,y\}\), let
\(\boldsymbol{\mathcal{F}}^d_{g,f,q}\in\mathbb{R}^{N_Q}\) denote the nodal
flux vector in Eq.~\eqref{eq:UGKSflux}. Its coefficient vector
is
\begin{equation}
 \boldsymbol{\mathcal{F}}_{g,f,q}^{\mathrm{M},d}
 =
 \Psi^{\mathsf T}W
 \boldsymbol{\mathcal{F}}^d_{g,f,q}.
 \label{eq:modalFluxProjection}
\end{equation}
The superscript \(\mathrm M\) denotes evaluation from basis coefficients.
With \(\partial_tB_{g,f}=0\), substituting
Eq.~\eqref{eq:modalInterfaceIntensity} into the time-averaged flux
definition~\eqref{eq:nodalMicroscopicFluxDefinition} gives the
flux coefficient vectors
\begin{align}
 \boldsymbol{\mathcal{F}}_{g,f,q}^{\mathrm{M},x}
 ={}&
 \mathcal C_1s_x^{(q)}B_{g,f}J^\eta\boldsymbol{e}_0
 +\mathcal C_2\left[
 (\partial_xB_{g,f})K^{\eta\eta}\boldsymbol{e}_0
 +s_x^{(q)}s_y^{(q)}(\partial_yB_{g,f})K^{\eta\chi}\boldsymbol{e}_0
 \right]
 \nonumber\\
 &+\mathcal C_3s_x^{(q)}J^\eta\boldsymbol{a}_{f,g,q}^{\mathrm{up},n}
 +\mathcal C_4\left[
 K^{\eta\eta}\boldsymbol{a}_{x,f,g,q}^{\mathrm{up},n}
 +s_x^{(q)}s_y^{(q)}K^{\eta\chi}\boldsymbol{a}_{y,f,g,q}^{\mathrm{up},n}
 \right],
 \label{eq:modalXFaceFlux}\\
 \boldsymbol{\mathcal{F}}_{g,f,q}^{\mathrm{M},y}
 ={}&
 \mathcal C_1s_y^{(q)}B_{g,f}J^\chi\boldsymbol{e}_0
 +\mathcal C_2\left[
 s_x^{(q)}s_y^{(q)}(\partial_xB_{g,f})K^{\eta\chi}\boldsymbol{e}_0
 +(\partial_yB_{g,f})K^{\chi\chi}\boldsymbol{e}_0
 \right]
 \nonumber\\
 &+\mathcal C_3s_y^{(q)}J^\chi\boldsymbol{a}_{f,g,q}^{\mathrm{up},n}
 +\mathcal C_4\left[
 s_x^{(q)}s_y^{(q)}K^{\eta\chi}\boldsymbol{a}_{x,f,g,q}^{\mathrm{up},n}
 +K^{\chi\chi}\boldsymbol{a}_{y,f,g,q}^{\mathrm{up},n}
 \right].
 \label{eq:modalYFaceFlux}
\end{align}
Equations~\eqref{eq:modalXFaceFlux} and \eqref{eq:modalYFaceFlux} give the
basis representation of the UGKS flux. Let
\(d_f\in\{x,y\}\) denote the normal coordinate direction of face \(\Gamma_f\).
Applying \(\Psi^{\mathsf T}W\) to Eq.~\eqref{eq:implicitNodalUpdate} in each quadrant
gives the corresponding microscopic update
\begin{equation}
 \begin{aligned}
 \left(1+\lambda_{i,g}^{n+1}\right)
 \widetilde{\boldsymbol a}_{i,g,q}^{\,n+1}
 ={}&\boldsymbol a_{i,g,q}^{n}
 -\frac{\Delta t}{|V_i|}
 \sum_{\Gamma_f\subset\partial V_i}
 s_{i,f}|\Gamma_f|\,
 \boldsymbol{\mathcal F}_{g,f,q}^{\mathrm M,d_f}\\
 &+\lambda_{i,g}^{n+1}B_{i,g}^{n+1}\boldsymbol e_0.
 \end{aligned}
 \label{eq:fullModalUGKSUpdate}
\end{equation}
The tilde denotes the coefficient vector before correction of its zeroth moment.

Expressing Eq.~\eqref{eq:frozenAngularMoments} in terms of the reconstructed
basis coefficients and their spatial slopes gives
\begin{equation}
 \begin{aligned}
  \mathcal G_{3,g,f}^{\mathrm M,n}
  ={}&\pi\sum_{q=0}^{3}\sum_{k=0}^{D_1-1}
  \bigl[
   n_{f,x}s_x^{(q)}(J^\eta)_{0k}
   +n_{f,y}s_y^{(q)}(J^\chi)_{0k}
  \bigr]
  a_{f,g,q,k}^{\mathrm{up},n},\\
  \mathcal G_{4,g,f}^{\mathrm M,n}
  ={}&\pi\sum_{q=0}^{3}\sum_{k=0}^{D_2-1}\Bigl[
  \bigl(
   n_{f,x}(K^{\eta\eta})_{0k}
   +n_{f,y}s_x^{(q)}s_y^{(q)}(K^{\eta\chi})_{0k}
  \bigr)
  a_{x,f,g,q,k}^{\mathrm{up},n}
  \\
  &\hspace{5.2em}+
  \bigl(
   n_{f,x}s_x^{(q)}s_y^{(q)}(K^{\eta\chi})_{0k}
   +n_{f,y}(K^{\chi\chi})_{0k}
  \bigr)
  a_{y,f,g,q,k}^{\mathrm{up},n}
  \Bigr].
 \end{aligned}
 \label{eq:adaptiveFrozenAngularMoments}
\end{equation}
Here \(n_{f,x}\) and \(n_{f,y}\) are the components of \(\boldsymbol n_f\).
Only the zeroth matrix rows are used; the bounds \(D_1\) and \(D_2\) follow
from Proposition~\ref{prop:modalBlockStructure} and its quadratic analogue.
These moments remain fixed during the coupled iteration in
Section~\ref{sec:coupledUGKSUpdate}.
At iteration \(\ell\), the macroscopic flux is
\begin{equation}
 \mathcal F_{g,f}^{\mathrm M,\rho,(\ell)}
 =\frac{4\pi}{3}\mathcal C_{2,f,g}^{(\ell)}
  \boldsymbol n_f\cdot\nabla B_{g,f}^{(\ell)}
 +\mathcal C_{3,f,g}^{(\ell)}\mathcal G_{3,g,f}^{\mathrm M,n}
 +\mathcal C_{4,f,g}^{(\ell)}\mathcal G_{4,g,f}^{\mathrm M,n}.
 \label{eq:adaptivePicardMacroscopicFlux}
\end{equation}
Evaluate \(A_i^{(\ell)}\) and \(\mathcal D_i^{(\ell)}\) from
Eq.~\eqref{eq:coupledTemperatureCoefficients} using this flux.
The coupled Picard update is
\begin{equation}
 \left\{
 \begin{aligned}
 &\begin{aligned}
   \left(1+\lambda_{i,g}^{(\ell)}\right)\rho_{i,g}^{(\ell+1)}
   ={}&\rho_{i,g}^{n}
   -\frac{\Delta t}{|V_i|}
   \sum_{\Gamma_f\subset\partial V_i}
   s_{i,f}|\Gamma_f|\,\mathcal F_{g,f}^{\mathrm M,\rho,(\ell)}\\
   &+ac\lambda_{i,g}^{(\ell)}\mathcal P_{i,g}^{(\ell)}
   \left(T_i^{(\ell+1)}\right)^4,
  \end{aligned}\\[6pt]
 &U_m\!\left(T_i^{(\ell+1)}\right)
  +ac\mathcal D_i^{(\ell)}\left(T_i^{(\ell+1)}\right)^4
  =U_m(T_i^n)+A_i^{(\ell)}.
 \end{aligned}
 \right.
 \label{eq:adaptivePicardCoupledUpdate}
\end{equation}
After convergence, evaluate Eq.~\eqref{eq:fullModalUGKSUpdate} using
\(T_i^{n+1}\). To ensure consistency between the microscopic and macroscopic
updates, we correct the constant coefficients:
\[
 \boldsymbol a_{i,g,q}^{n+1}
 =\widetilde{\boldsymbol a}_{i,g,q}^{\,n+1}
 +\frac{\displaystyle\rho_{i,g}^{n+1}
  -\pi\sum_{q'=0}^{3}\widetilde a_{i,g,q',0}^{\,n+1}}{4\pi}
 \boldsymbol e_0.
\]
Thus
\[
 \pi\sum_{q=0}^{3}a_{i,g,q,0}^{n+1}=\rho_{i,g}^{n+1},
 \qquad
 a_{i,g,q,k}^{n+1}=\widetilde a_{i,g,q,k}^{\,n+1},\quad 1\le k<N_Q.
\]

\section{Angular DoF adaptation}
\label{sec:adaptation}

This section presents the adaptive strategy for angular degrees of freedom.
For fixed \((i,g,q)\), let \(p=p_{i,g,q}^{n}\) denote the local truncation degree.
The radiation intensity distribution function is represented by
\begin{equation}
 \boldsymbol I_{i,g,q}^{n}
 =\sum_{k=0}^{D_p-1}a_{i,g,q,k}^{n}\boldsymbol\psi_k,
 \qquad
 \boldsymbol a_{i,g,q}^{n}\in\mathbb R^{D_p},
 \label{eq:localStoredModalVector}
\end{equation}
The local number of angular DoFs is \(D_p\), as defined in
Section~\ref{sec:modalBasis}. Define
\begin{equation}
 a_{i,g,q,k}^{n}=0,
 \qquad
 D_p\leq k<N_Q.
 \label{eq:vanishingHigherCoefficients}
\end{equation}

The adaptive update involves several steps, including candidate space construction,
evolution of basis coefficients, energy consistency correction, and coefficient truncation.

\subsection{Construction of the local candidate space}
The basis coefficients to be updated are those retained from the previous
time step after exponential decay and those required by interface transport.

\smallskip
\noindent\textit{(i) Surviving nonequilibrium contribution.}
Motivated by the UGKWP survival probability, we select the surviving truncation degree
using the decay factor \(e^{-\lambda_{i,g}^{n+1}}\).
For \(0\leq r\leq p_{i,g,q}^{n}\), define the norm of the coefficients above
degree \(r\) by
\begin{equation}
 \mathcal T_{i,g,q}^{n}(r)
 =
 \left[
  \sum_{k=D_r}^{D_{p_{i,g,q}^{n}}-1}
  \bigl(a_{i,g,q,k}^{n}\bigr)^2
 \right]^{1/2},
 \qquad
 \mathcal T_{i,g,q}^{n}(p_{i,g,q}^{n})=0.
 \label{eq:stateModalTail}
\end{equation}
Here \(D_p\) is the number of basis vectors with degree at most \(p\).
For the relative tolerance \(\varepsilon_{\mathrm{surv}}\), define the surviving truncation degree by
\begin{equation}
 p_{i,g,q}^{\mathrm{surv}}
 =
 \min\left\{
 r\in\{0,\ldots,p_{i,g,q}^{n}\}
 \;\middle|\;
 e^{-\lambda_{i,g}^{n+1}}\mathcal T_{i,g,q}^{n}(r)
 \leq
 \varepsilon_{\mathrm{surv}}
 \|\boldsymbol a_{i,g,q}^{n}\|_2
 \right\}.
 \label{eq:survivingOrder}
\end{equation}

\smallskip
\noindent\textit{(ii) Interface transport.}
To select the flux truncation degree without evaluating all basis coefficients
of the interface flux, we bound the truncation error using the tails of the
reconstructed coefficients and their slopes. Using the upwind coefficients from
Section~\ref{sec:modalUGKS}, define
\begin{equation}
 \mathcal T_{f,g,q}(r)
 =
 \left[
  \sum_{k=D_r}^{D_{p_{f,g,q}^{\mathrm{up}}}-1}
  \bigl(a_{f,g,q,k}^{\mathrm{up},n}\bigr)^2
 \right]^{1/2},
 \label{eq:reconstructedModalTails}
\end{equation}
where \(p_{f,g,q}^{\mathrm{up}}\) is the local truncation degree in the upwind cell.
For the \(x\) derivative, define
\begin{equation}
 \mathcal T_{f,g,q}^{x}(r)
 =
 \left[
  \sum_{k=D_r}^{D_{p_{f,g,q}^{\mathrm{up}}}-1}
  \bigl(a_{x,f,g,q,k}^{\mathrm{up},n}\bigr)^2
 \right]^{1/2}.
 \label{eq:reconstructedSlopeTails}
\end{equation}
The \(y\) derivative gives \(\mathcal T_{f,g,q}^{y}(r)\) analogously.
All three tail norms are zero for \(r\geq p_{f,g,q}^{\mathrm{up}}\).

For \(p_{\mathrm{full}}\geq2\), Proposition~\ref{prop:modalBlockStructure} and
its quadratic analogue give the upper bound on the flux expansion degree:
\begin{equation}
 \overline p_{f,g,q}
 =\min\{p_{\mathrm{full}},p_{f,g,q}^{\mathrm{up}}+2\}.
 \label{eq:maximalFaceOutputOrder}
\end{equation}

For the \(x\) direction, the flux coefficient vector is
\begin{equation}
 \boldsymbol{\mathcal F}_{g,f,q,\overline p_{f,g,q}}^{\mathrm M,x}
 =
 \left[
  \mathcal F_{g,f,q,0}^{\mathrm M,x},\ldots,
  \mathcal F_{g,f,q,D_{\overline p_{f,g,q}}-1}^{\mathrm M,x}
 \right]^{\mathsf T}
 \in\mathbb R^{D_{\overline p_{f,g,q}}}.
 \label{eq:referenceAndTrialFluxes}
\end{equation}
The equilibrium flux terms have degree at most two and are retained in full
by taking \(r\geq2\). Truncation at degree
\(2\leq r\leq\overline p_{f,g,q}\) retains
\begin{equation}
 \boldsymbol{\mathcal F}_{g,f,q,r}^{\mathrm M,x}
 =
 \left[
  \mathcal F_{g,f,q,0}^{\mathrm M,x},\ldots,
  \mathcal F_{g,f,q,D_r-1}^{\mathrm M,x}
 \right]^{\mathsf T}
 \in\mathbb R^{D_r}.
 \label{eq:truncatedFaceFlux}
\end{equation}
By orthonormality, the flux truncation error has norm
\begin{equation}
 \mathcal E_{f,g,q}^{x}(r)
 =
 \left[
  \sum_{k=D_r}^{D_{\overline p_{f,g,q}}-1}
  \bigl(\mathcal F_{g,f,q,k}^{\mathrm M,x}\bigr)^2
 \right]^{1/2}.
 \label{eq:faceFluxTruncationDifference}
\end{equation}
The \(y\) flux and its truncation error \(\mathcal E_{f,g,q}^{y}(r)\) are
defined analogously.

\begin{proposition}[Flux coefficient truncation estimate]
\label{prop:modalFluxTruncationEstimate}
For each interface \(\Gamma_f\), frequency group \(g\), quadrant \(q\), and
\(2\leq r\leq\overline p_{f,g,q}\),
\begin{equation}
 \max\left\{
  \mathcal E_{f,g,q}^{x}(r),
  \mathcal E_{f,g,q}^{y}(r)
 \right\}
 \leq
 |\mathcal C_3|\mathcal T_{f,g,q}(r-1)
 +|\mathcal C_4|\left[
  \mathcal T_{f,g,q}^{x}(r-2)
  +\mathcal T_{f,g,q}^{y}(r-2)
 \right].
 \label{eq:discardedFluxBound}
\end{equation}
\end{proposition}

\begin{proof}
Orthonormality of the full basis and \(0\leq\eta_n\leq1\) give, for any
\(\boldsymbol v\in\mathbb R^{N_Q}\),
\begin{equation}
 \|J^\eta\boldsymbol v\|_2
 =\|E\Psi\boldsymbol v\|_W
 \leq\|\Psi\boldsymbol v\|_W
 =\|\boldsymbol v\|_2.
 \label{eq:modalOperatorContraction}
\end{equation}
The same bound holds for \(J^\chi\), \(K^{\eta\eta}\), \(K^{\eta\chi}\), and
\(K^{\chi\chi}\), since \(C\), \(E^2\), \(EC\), and \(C^2\) also have
diagonal entries in \([0,1]\).

For \(r\geq2\), the equilibrium terms vanish above degree \(r\).
By Proposition~\ref{prop:modalBlockStructure} and its quadratic analogue, linear
and quadratic angular multiplication increase the degree by at most one and two,
respectively. Thus flux coefficients above degree \(r\) depend only on coefficient
tails above degrees \(r-1\) and \(r-2\), respectively. For
\(D_r\leq\ell<D_{\overline p_{f,g,q}}\),
\begin{equation}
 \begin{aligned}
  \mathcal F_{g,f,q,\ell}^{\mathrm M,x}
  ={}&\mathcal C_3s_x^{(q)}
  \sum_{k=D_{r-1}}^{D_{p_{f,g,q}^{\mathrm{up}}}-1}
  (J^\eta)_{\ell k}a_{f,g,q,k}^{\mathrm{up},n}
  +\mathcal C_4
  \sum_{k=D_{r-2}}^{D_{p_{f,g,q}^{\mathrm{up}}}-1}
  (K^{\eta\eta})_{\ell k}a_{x,f,g,q,k}^{\mathrm{up},n}
  \\
  &+\mathcal C_4s_x^{(q)}s_y^{(q)}
  \sum_{k=D_{r-2}}^{D_{p_{f,g,q}^{\mathrm{up}}}-1}
  (K^{\eta\chi})_{\ell k}a_{y,f,g,q,k}^{\mathrm{up},n}.
 \end{aligned}
 \label{eq:highDegreeFluxCoefficients}
\end{equation}
Applying the triangle inequality and Eq.~\eqref{eq:modalOperatorContraction}
to the Euclidean norm of these coefficients gives
\begin{equation}
 \mathcal E_{f,g,q}^{x}(r)
 \leq
 |\mathcal C_3|\mathcal T_{f,g,q}(r-1)
 +|\mathcal C_4|\left[
  \mathcal T_{f,g,q}^{x}(r-2)
  +\mathcal T_{f,g,q}^{y}(r-2)
 \right].
 \label{eq:separateFluxTailBoundsInProof}
\end{equation}
The same bound holds for \(\mathcal E_{f,g,q}^{y}(r)\), which proves
Eq.~\eqref{eq:discardedFluxBound}.
\end{proof}

Let \(\mathcal N(f)\) denote the cells adjacent to \(\Gamma_f\). Truncation
of the interface flux at degree \(r\) changes the update of
\(V_i\in\mathcal N(f)\) by at most
\begin{equation}
 \mathcal E_{i,f,g,q}^{\mathrm{upd}}(r)
 =
 \frac{\Delta t\,|\Gamma_f|}{|V_i|}
 \left[
  |\mathcal C_3|\mathcal T_{f,g,q}(r-1)
  +|\mathcal C_4|\left(
   \mathcal T_{f,g,q}^{x}(r-2)
   +\mathcal T_{f,g,q}^{y}(r-2)
  \right)
 \right].
 \label{eq:tailindicator}
\end{equation}

For tolerance \(\varepsilon_{\mathrm{flux}}>0\), define the flux truncation degree by
\begin{equation}
 p_{f,g,q}^{\mathrm{out}}
 =
 \min\left(
  \left\{
   r\in\{2,\ldots,\overline p_{f,g,q}\}
   \;\middle|\;
    \max_{i\in\mathcal N(f)}\mathcal E_{i,f,g,q}^{\mathrm{upd}}(r)
    \leq\varepsilon_{\mathrm{flux}}
  \right\}
  \cup\{\overline p_{f,g,q}\}
 \right).
 \label{eq:adaptiveFaceOrder}
\end{equation}
This degree is selected before the Picard iteration and remains fixed
throughout the time step.

\smallskip
\noindent\textit{(iii) Local candidate space.}
Combining the surviving and interface contributions, define the candidate
truncation degree by
\begin{equation}
 p_{i,g,q}^{\mathrm c}
 =
 \max\left(
  p_{i,g,q}^{\mathrm{surv}},
  \max_{\Gamma_f\subset\partial V_i}p_{f,g,q}^{\mathrm{out}}
 \right).
 \label{eq:candidateOrder}
\end{equation}
Using Eqs.~\eqref{eq:modalXFaceFlux}--\eqref{eq:modalYFaceFlux}, we compute only
the first \(D_{p_{f,g,q}^{\mathrm{out}}}\) basis coefficients of the interface flux.

\subsection{Candidate space update and coefficient truncation}
\label{sec:candidateUpdate}

Suppressing \(i,g,q\), the candidate coefficient vector has
\(D_{p^{\mathrm c}}\) angular DoFs:
\[
 \boldsymbol a_{\mathrm c}^{n}
 =\bigl(a_0^n,\ldots,a_{D_{p^{\mathrm c}}-1}^{n}\bigr)^{\mathsf T}.
\]
It is formed by truncating or zero-extending the stored coefficient vector to
this size, with zero extension as in Eq.~\eqref{eq:vanishingHigherCoefficients}.
The interface flux is extended to the same space by
\begin{equation}
 \bigl(\boldsymbol{\mathcal F}_{f}^{\mathrm c}\bigr)_k
 =
 \begin{cases}
  \mathcal F_{f,k}^{\mathrm M,d_f},
  &0\leq k<D_{p_f^{\mathrm{out}}},\\
  0,
  &D_{p_f^{\mathrm{out}}}\leq k<D_{p^{\mathrm c}}.
 \end{cases}
 \label{eq:candidateFaceFluxEmbedding}
\end{equation}
Equation~\eqref{eq:fullModalUGKSUpdate} then becomes
\begin{equation}
 \bigl(1+\lambda^{n+1}\bigr)
 \widetilde{\boldsymbol a}^{\,n+1}
 =
 \boldsymbol a_{\mathrm c}^{n}
 -\frac{\Delta t}{|V|}
  \sum_{\Gamma_f\subset\partial V}
  s_f|\Gamma_f|
  \boldsymbol{\mathcal F}_{f}^{\mathrm c}
 +\lambda^{n+1}B^{n+1}\boldsymbol e_0.
 \label{eq:modalUGKSUpdate}
\end{equation}
Here \(\boldsymbol e_0=(1,0,\ldots,0)^{\mathsf T}\in\mathbb R^{D_{p^{\mathrm c}}}\).
For consistency with the macroscopic density, set
\begin{equation}
 \widehat a_{i,g,q,k}^{\,n+1}
 =\widetilde a_{i,g,q,k}^{\,n+1}
 +\frac{\delta_{k0}}{4\pi}\left(
  \rho_{i,g}^{n+1}
  -\pi\sum_{q'=0}^{3}\widetilde a_{i,g,q',0}^{\,n+1}
 \right).
 \label{eq:zeroMomentCorrection}
\end{equation}
Here \(\delta_{k0}\) is the Kronecker delta, \(q=0,\ldots,3\), and
\(0\leq k<D_{p_{i,g,q}^{\mathrm c}}\).

Define the coefficient tail norm by
\begin{equation}
 \mathcal T_{i,g,q}^{n+1}(r)
 =
 \left[
  \sum_{k=D_r}^{D_{p_{i,g,q}^{\mathrm c}}-1}
  \bigl(\widehat a_{i,g,q,k}^{\,n+1}\bigr)^2
 \right]^{1/2},
 \qquad 0\leq r\leq p_{i,g,q}^{\mathrm c}.
 \label{eq:updatedStateModalTail}
\end{equation}
For relative tolerance \(\varepsilon_{\mathrm{state}}>0\), define the truncation degree by
\begin{equation}
 p_{i,g,q}^{n+1}
 =
 \min\left\{
 r\in\{0,\ldots,p_{i,g,q}^{\mathrm c}\}
 \;\middle|\;
 \mathcal T_{i,g,q}^{n+1}(r)
 \leq
 \frac{\varepsilon_{\mathrm{state}}}{2}
 \left(
 \sum_{q'=0}^{3}
  \bigl\|\widetilde{\boldsymbol a}_{i,g,q'}^{\,n+1}\bigr\|_2^2
 \right)^{1/2}
 \right\}.
 \label{eq:retainedOrder}
\end{equation}
The reference norm combines the pre-correction coefficient vectors in all four
quadrants. The factor \(1/2\) allocates an equal share of the squared error budget
to each quadrant.
The retained coefficient vector is
\begin{equation}
 \boldsymbol a_{i,g,q}^{n+1}
 =\left(
  \widehat a_{i,g,q,0}^{\,n+1},\ldots,
  \widehat a_{i,g,q,D_{p_{i,g,q}^{n+1}}-1}^{\,n+1}
 \right)^{\mathsf T}
 \in\mathbb R^{D_{p_{i,g,q}^{n+1}}}.
 \label{eq:storedUpdatedModalVector}
\end{equation}
By Eq.~\eqref{eq:modalParseval}, the combined squared \(W\)-norm projection error satisfies
\[
 \sum_{q=0}^{3}
 \left[\mathcal T_{i,g,q}^{n+1}\!\left(p_{i,g,q}^{n+1}\right)\right]^2
 \leq\varepsilon_{\mathrm{state}}^2
 \sum_{q=0}^{3}
 \bigl\|\widetilde{\boldsymbol a}_{i,g,q}^{\,n+1}\bigr\|_2^2.
\]

\subsection{Coupling with the material temperature equation}

The Picard iteration uses Eqs.~\eqref{eq:adaptivePicardMacroscopicFlux}
and~\eqref{eq:adaptivePicardCoupledUpdate}, with the moments
\(\mathcal G_{3,g,f}^{\mathrm M,n}\) and
\(\mathcal G_{4,g,f}^{\mathrm M,n}\) from
Eq.~\eqref{eq:adaptiveFrozenAngularMoments} held fixed.

After convergence, apply the update, correction, and truncation in
Section~\ref{sec:candidateUpdate} using \(T^{n+1}\) and \(\rho^{n+1}\).
Algorithm~\ref{alg:angularAdaptation} summarizes one time step.
\begin{algorithm}[H]
\caption{One coupled time step of NCUGKS}
\label{alg:angularAdaptation}
\begin{algorithmic}[1]
 \State \textbf{Given}\quad
 \(\{p_{i,g,q}^{n},\boldsymbol a_{i,g,q}^{n}\}\),
 \(\{\rho_{i,g}^{n}\}\), \(\{T_i^n\}\), \(\Delta t\),
 \(\varepsilon_{\mathrm{surv}}\), \(\varepsilon_{\mathrm{flux}}\),
 \(\varepsilon_{\mathrm{state}}\), \(\varepsilon_T\), and
 \(\varepsilon_\rho\)
 \State Reconstruct the time-\(n\) basis coefficients and slopes in physical
 space.
 \State Form \(\mathcal G_{3,g,f}^{\mathrm M,n}\) and \(\mathcal G_{4,g,f}^{\mathrm M,n}\) from
 Eq.~\eqref{eq:adaptiveFrozenAngularMoments}, and the tail norms
 \(\mathcal T_{f,g,q}(r)\), \(\mathcal T_{f,g,q}^{x}(r)\), and
 \(\mathcal T_{f,g,q}^{y}(r)\) from
 Eqs.~\eqref{eq:reconstructedModalTails}--\eqref{eq:reconstructedSlopeTails}.
 \State Determine \(p_{f,g,q}^{\mathrm{out}}\) from
 Eq.~\eqref{eq:adaptiveFaceOrder} using the time-\(n\) face coefficients.
 \State Set \(T_i^{(0)}=T_i^n\),
 \(\rho_{i,g}^{(0)}=\rho_{i,g}^n\), and \(\ell=0\).
 \Repeat
  \State Evaluate the cell-centered quantities in
  Eq.~\eqref{eq:collisionParameters} and \(B_{i,g}^{(\ell)}\) at
  \(T_i^{(\ell)}\).
  \State Form \(B_{g,f}^{(\ell)}\), \(\nabla B_{g,f}^{(\ell)}\), and
  \(\sigma_{g,f}^{(\ell)}\) using
  Eqs.~\eqref{eq:faceEquilibriumOpacity} and
  \eqref{eq:faceEquilibriumGradient}.
  \State Evaluate \(\mathcal C_{j,f,g}^{(\ell)}\), \(j=1,\ldots,4\), from
  Eqs.~\eqref{eq:q12}--\eqref{eq:q34}.
  \State Form \(\mathcal F_{g,f}^{\mathrm M,\rho,(\ell)}\) from
  Eq.~\eqref{eq:adaptivePicardMacroscopicFlux}.
  \State Form \(A_i^{(\ell)}\) and \(\mathcal D_i^{(\ell)}\) from
  Eq.~\eqref{eq:coupledTemperatureCoefficients}, using
  \(\mathcal F_{g,f}^{\mathrm M,\rho,(\ell)}\) in \(A_i^{(\ell)}\).
  \State Solve Eq.~\eqref{eq:adaptivePicardCoupledUpdate} for
  \(T_i^{(\ell+1)}\) and \(\rho_{i,g}^{(\ell+1)}\).
  \State \(\ell\gets\ell+1\).
 \Until{\(\max_i|T_i^{(\ell)}-T_i^{(\ell-1)}|\leq\varepsilon_T\) and
 \(\max_{i,g}|\rho_{i,g}^{(\ell)}-\rho_{i,g}^{(\ell-1)}|
 \leq\varepsilon_\rho\)}
 \State Accept \(T_i^{n+1}=T_i^{(\ell)}\),
 \(\rho_{i,g}^{n+1}=\rho_{i,g}^{(\ell)}\), and the macroscopic flux
 \(\mathcal F_{g,f}^{\mathrm M,\rho,(\ell-1)}\) of the final Picard sweep.
 \State Refresh the cell and face coefficients from \(T_i^{n+1}\).
 \State Evaluate \(\lambda_{i,g}^{n+1}\) and determine
 \(p_{i,g,q}^{\mathrm{surv}}\) from Eq.~\eqref{eq:survivingOrder}.
 \State Evaluate the interface flux coefficients from
 Eqs.~\eqref{eq:modalXFaceFlux}--\eqref{eq:modalYFaceFlux} for
 \(0\leq k<D_{p_{f,g,q}^{\mathrm{out}}}\).
 \State Determine the candidate degree \(p_{i,g,q}^{\mathrm c}\) from
 Eq.~\eqref{eq:candidateOrder}.
 \State Advance Eq.~\eqref{eq:modalUGKSUpdate} in the candidate space.
 \State Apply the consistency correction in
 Eq.~\eqref{eq:zeroMomentCorrection}.
 \State Select \(p_{i,g,q}^{n+1}\) from Eq.~\eqref{eq:retainedOrder} and store
 the retained basis coefficients.
\end{algorithmic}
\end{algorithm}

\subsection{Optically thick limit of the angular DoFs}
\label{sec:angularDofAsymptotics}

With the mesh, \(\Delta t>0\), and angular quadrature fixed, consider
\[
 \lambda_{i,g}^{n+1}\to\infty,
 \qquad \tau_{g,f}\to0.
\]
Equations~\eqref{eq:q12}--\eqref{eq:q34} give
\begin{equation}
 \mathcal C_{1,f,g}\to c,
 \qquad
 \mathcal C_{2,f,g},\ \mathcal C_{3,f,g},\ \mathcal C_{4,f,g}\to0.
 \label{eq:angularDofScales}
\end{equation}

\begin{samepage}
\begin{proposition}[Reduction of the angular DoFs]
\label{prop:angularDofThickLimit}
Let \(p_{\mathrm{full}}\geq2\) and suppose
\[
 \sum_{q'=0}^{3}
 \bigl\|\widetilde{\boldsymbol a}_{i,g,q'}^{\,n+1}\bigr\|_2^2
 \geq C_*>0,
\]
where \(C_*\) is independent of opacity.
For sufficiently large opacities \(\sigma_g\),
\begin{equation}
 \left(
 p_{i,g,q}^{\mathrm{surv}},
 p_{f,g,q}^{\mathrm{out}},
 p_{i,g,q}^{\mathrm c},
 p_{i,g,q}^{n+1}
 \right)
 =(0,2,2,0).
 \label{eq:angularDofLimitingOrders}
\end{equation}
Thus each quadrant retains only its constant basis coefficient, giving four
angular degrees of freedom per cell and frequency group.
\end{proposition}
\end{samepage}

\begin{proof}
For sufficiently large \(\lambda_{i,g}^{n+1}\),
\[
 e^{-\lambda_{i,g}^{n+1}}\mathcal T_{i,g,q}^{n}(0)
 \leq e^{-\lambda_{i,g}^{n+1}}\|\boldsymbol a_{i,g,q}^{n}\|_2
 \leq\varepsilon_{\mathrm{surv}}\|\boldsymbol a_{i,g,q}^{n}\|_2.
\]
Equation~\eqref{eq:survivingOrder} therefore gives
\(p_{i,g,q}^{\mathrm{surv}}=0\).
Proposition~\ref{prop:modalFluxTruncationEstimate} gives
\begin{equation}
 \max\{\mathcal E_{f,g,q}^{x}(2),\mathcal E_{f,g,q}^{y}(2)\}
 \leq
 |\mathcal C_3|\mathcal T_{f,g,q}(1)
 +|\mathcal C_4|\left[
  \mathcal T_{f,g,q}^{x}(0)+\mathcal T_{f,g,q}^{y}(0)
 \right]
 \longrightarrow0.
 \label{eq:angularDofFluxEstimate}
\end{equation}
Since \(\max_{i\in\mathcal N(f)}\mathcal E_{i,f,g,q}^{\mathrm{upd}}(2)\to0\),
Eqs.~\eqref{eq:tailindicator}--\eqref{eq:candidateOrder} give
\[
 p_{f,g,q}^{\mathrm{out}}=2,
 \qquad p_{i,g,q}^{\mathrm c}=2
\]
for sufficiently large opacities.

For \(1\leq k<D_2\), Eq.~\eqref{eq:modalUGKSUpdate} yields
\begin{equation}
 \widetilde a_{i,g,q,k}^{\,n+1}
 =\frac{1}{1+\lambda_{i,g}^{n+1}}
 \left[
  a_{i,g,q,k}^{n}
  -\frac{\Delta t}{|V_i|}
   \sum_{\Gamma_f\subset\partial V_i}
   s_{i,f}|\Gamma_f|
   \mathcal F_{g,f,q,k}^{\mathrm M,d_f}
 \right]
 \longrightarrow0,
 \label{eq:angularDofNonconstantUpdate}
\end{equation}
since the bracketed terms are bounded.
The correction in Eq.~\eqref{eq:zeroMomentCorrection} leaves these coefficients unchanged, so
\[
 \mathcal T_{i,g,q}^{n+1}(0)
 =\left[\sum_{k=1}^{D_2-1}
   \bigl(\widetilde a_{i,g,q,k}^{\,n+1}\bigr)^2\right]^{1/2}
 \longrightarrow0.
\]
The right-hand side of Eq.~\eqref{eq:retainedOrder} is at least
\(\varepsilon_{\mathrm{state}}\sqrt{C_*}/2>0\).
Since \(\mathcal T_{i,g,q}^{n+1}(0)\to0\), \(r=0\) satisfies
Eq.~\eqref{eq:retainedOrder} for sufficiently large opacities, giving
\(p_{i,g,q}^{n+1}=0\).
\end{proof}

\begin{remark}
\label{rem:isotropicStorage}
For an isotropic state, \(a_{i,g,q,0}^{n+1}=\rho_{i,g}^{n+1}/(4\pi)\),
\(q=0,\ldots,3\). These coefficients can therefore be recovered from the group
density without separate storage or any change to the basis representation.
\end{remark}

\section{Numerical examples}
\label{sec:numerics}

To assess the accuracy, angular DoF adaptation, and computational efficiency
of the present method, we consider numerical tests spanning different optical
regimes. The results are compared with those of
other methods.

All test cases use a \(48\times48\) product angular quadrature.
Let \(\overline D_{\mathrm{ang}}(t)\) denote the number of separately stored angular
coefficients at time \(t\), averaged over all cells and frequency groups.
States represented isotropically by the group density contribute zero to this mean,
as described in Remark~\ref{rem:isotropicStorage}.
The retained degree \(p_{i,g,q}(t)\) is defined in
Section~\ref{sec:adaptation}, with
\(p_{\max}(t)=\max_{i,g,q}p_{i,g,q}(t)\).
The retained fraction is
\[
 f_{\mathrm{ang}}(t)
 =\frac{\overline D_{\mathrm{ang}}(t)}{(48\times48)/2}.
\]
The denominator is halved because the intensities at \((\mu,\theta)\) and
\((-\mu,\theta)\) coincide, as described in Section~\ref{sec:discreteAngularSpace}.

Computational cost and memory are compared between NCUGKS and the original UGKS,
both using a single CPU core. The speedup in mean time per step is
\[
 S=\frac{\overline t_{\mathrm{UGKS}}}{\overline t_{\mathrm{NCUGKS}}}.
\]

{\emergencystretch=2em
Memory is measured in MiB. The ratio is
\(M^{\mathrm{UGKS}}/M^{\mathrm{NCUGKS}}\).
The corresponding reduction is
\mbox{\(100(1-M^{\mathrm{NCUGKS}}/M^{\mathrm{UGKS}})\%\)}.\par}

\subsection{Homogeneous multigroup Marshak wave}
\label{sec:homogeneousMarshak}

The homogeneous frequency-dependent Marshak problem~\cite{SunJiangXuLi2015}
is tested. The physical domain is
\(\left[0,5\right]\times\left[0,0.01\right]\,\mathrm{cm}^{2}\). The opacity is
\begin{equation}
 \sigma_{\nu}(T)
 =
 \frac{\sigma_0}{(h_{\mathrm P}\nu)^3\sqrt{T}},
 \qquad
 \sigma_0\in\{10,100,1000\}\
 \mathrm{keV}^{7/2}\mathrm{cm}^{-1}.
 \label{eq:homogeneousMarshakOpacity}
\end{equation}
Initially, \(T_m=T_r=10^{-3}\,\mathrm{keV}\). A half-range Planck inflow at
\(1\,\mathrm{keV}\) is imposed at \(x=0\); the other boundaries are
reflecting. The material energy is \(U_m=C_vT_m\), with
\(C_v=0.1\,\mathrm{GJ}\,\mathrm{cm}^{-3}\mathrm{keV}^{-1}\). The radiation
temperature is
\[
 T_r
 =
 \left(\frac{\sum_{g=1}^{G}\rho_g}{ac}\right)^{1/4}.
\]

The spatial mesh has \(1000\times2\) uniform cells. The frequency interval
corresponding to photon energies
\(\epsilon=h_{\mathrm P}\nu\in[10^{-4},10^{2}]\,\mathrm{keV}\)
is divided into 24 logarithmically spaced groups. The
time step is \(8.3389\times10^{-5}\,\mathrm{ns}\), and the solution is advanced
to \(t=1\,\mathrm{ns}\). The angular tolerances in
Eqs.~\eqref{eq:survivingOrder}, \eqref{eq:adaptiveFaceOrder}, and
\eqref{eq:retainedOrder} are
\mbox{\(\varepsilon_{\mathrm{surv}}=10^{-2}\)} and
\mbox{\(\varepsilon_{\mathrm{flux}}=\varepsilon_{\mathrm{state}}=10^{-4}\)}.

\begin{figure}[!htbp]
 \centering
 \includegraphics[width=\textwidth]{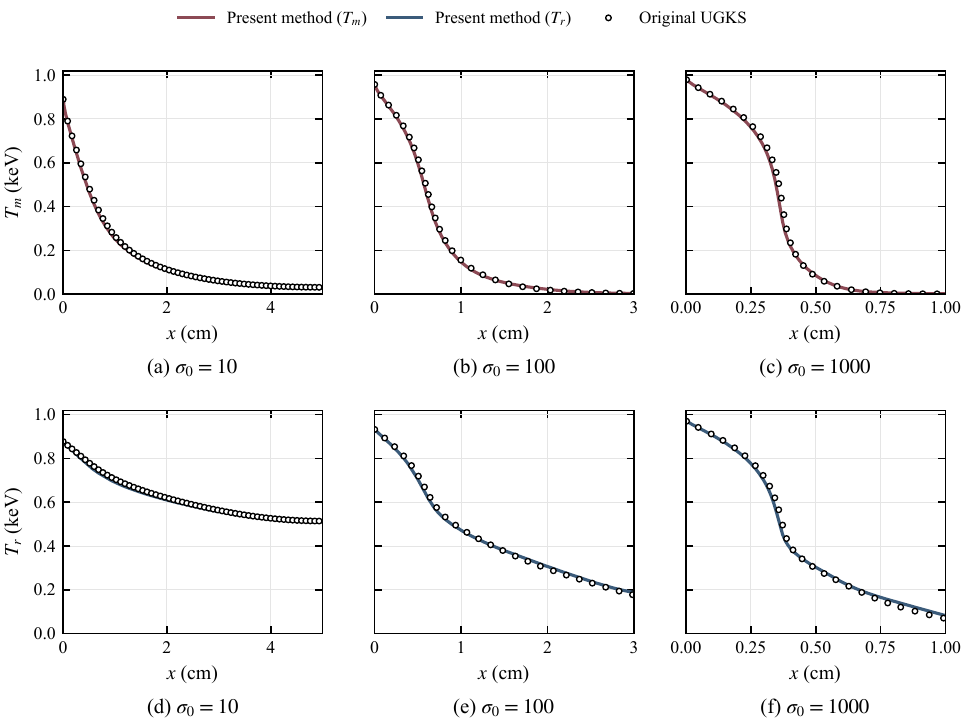}
 \caption{Homogeneous Marshak wave at \(t=1\,\mathrm{ns}\): material temperature
 (a--c) and radiation temperature (d--f) for the three opacity scales. Solid
 curves denote the present method; symbols denote the original UGKS results of
 Sun et al.~\cite{SunJiangXuLi2015}.}
 \label{fig:homogeneousMarshakProfiles}
\end{figure}

As shown in Fig.~\ref{fig:homogeneousMarshakProfiles}, the material and
radiation temperature results agree closely with the original UGKS results
for all three opacity scales.

\begin{figure}[!htbp]
 \centering
 \includegraphics[width=\textwidth]{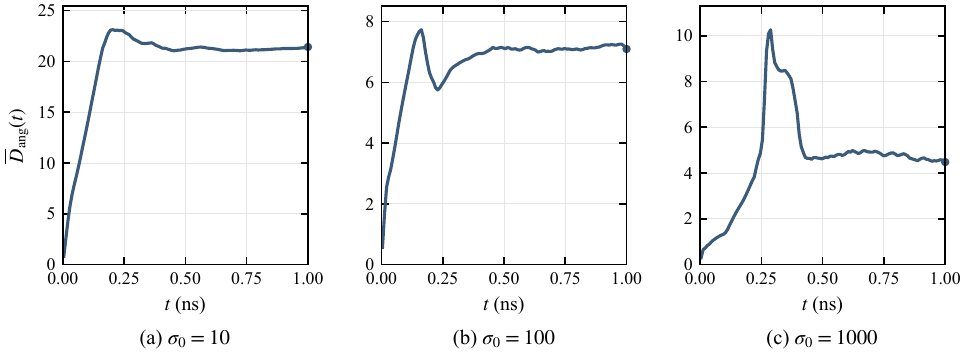}
 \caption{Mean retained DoFs for the homogeneous Marshak wave.
 Circles mark the final values.}
 \label{fig:homogeneousMarshakDofHistory}
\end{figure}

As summarized in Table~\ref{tab:homogeneousMarshakResults}, the final mean
number of retained DoFs decreases from \(21.42\) to \(4.48\) as
\(\sigma_0\) increases from \(10\) to \(1000\).
For \(\sigma_0=10\), it reaches \(23.12\) near
\(t=0.20\,\mathrm{ns}\) and then decreases slowly, as shown in
Fig.~\ref{fig:homogeneousMarshakDofHistory}. For \(\sigma_0=100\), it
reaches \(7.73\) near \(0.16\,\mathrm{ns}\), falls to \(5.75\) near
\(0.23\,\mathrm{ns}\), and subsequently increases to \(7.10\).
For \(\sigma_0=1000\), the maximum is \(10.26\) near
\(0.29\,\mathrm{ns}\), followed by a decrease to \(4.48\).
The small initial values of \(\overline D_{\mathrm{ang}}\) reflect that most states
are initially represented by their group densities alone, as described in
Remark~\ref{rem:isotropicStorage}.

\begin{table}[!htbp]
 \centering
 \caption{Retained DoFs for the homogeneous Marshak wave at
 \(t=1\,\mathrm{ns}\).}
 \label{tab:homogeneousMarshakResults}
 \normalsize
 \renewcommand{\arraystretch}{1.18}
 \begin{tabular*}{0.86\textwidth}{@{\hspace{8pt}\extracolsep{\fill}}cccc@{\hspace{8pt}}}
 \toprule
 \(\sigma_0\) & \(\overline D_{\mathrm{ang}}\) &
 \(f_{\mathrm{ang}}\) (\%) & \(p_{\max}\) \\
 \midrule
 10   & 21.42 & 1.860 & 14 \\
 100  &  7.10 & 0.616 &  9 \\
 1000 &  4.48 & 0.389 & 16 \\
 \bottomrule
 \end{tabular*}
\end{table}

Table~\ref{tab:homogeneousMarshakMemory} shows that the speedup increases
from \(56.3\) to \(257.9\) with opacity, consistent with the decreasing
number of retained DoFs. The memory requirement is reduced by
\(98.63\%\)--\(99.35\%\). These results illustrate the nonequilibrium-capturing
mechanism of NCUGKS. The angular DoFs adapt to the transport regime, with a lower
final mean at higher opacities.

\begin{table}[!htbp]
 \centering
 \caption{Computational cost and memory for the homogeneous Marshak wave.}
 \label{tab:homogeneousMarshakMemory}
 \small
 \setlength{\tabcolsep}{4pt}
 \begin{tabular}{rlrrrr}
 \toprule
 \(\sigma_0\) & Quantity & Original UGKS & Present method & Ratio & Reduction (\%) \\
 \midrule
 10 & Mean step time (s) & 2.937 & 0.05218 & 56.3 & --- \\
    & \(M\) (MiB) & 4236.33 & 57.91 & 73.15 & 98.63 \\
 \addlinespace
 100 & Mean step time (s) & 2.649 & 0.02078 & 127.5 & --- \\
     & \(M\) (MiB) & 4236.33 & 27.51 & 154.01 & 99.35 \\
 \addlinespace
 1000 & Mean step time (s) & 3.511 & 0.01361 & 257.9 & --- \\
      & \(M\) (MiB) & 4236.33 & 32.26 & 131.33 & 99.24 \\
 \bottomrule
 \end{tabular}
\end{table}

\FloatBarrier
\subsection{Heterogeneous Marshak wave}
\label{sec:heterogeneousMarshak}

We test the heterogeneous Marshak wave in the thin-to-thick and thick-to-thin
cases. The physical domains are
\([0,3]\times[0,0.01]\,\mathrm{cm}^{2}\) and
\([0,1.5]\times[0,0.01]\,\mathrm{cm}^{2}\), respectively.
The coefficient \(\sigma_0\) in Eq.~\eqref{eq:homogeneousMarshakOpacity} is
piecewise constant:
\[
 \begin{aligned}
 \sigma_0(x)
 &=
 \begin{cases}
  10,   & 0\leq x<2\,\mathrm{cm},\\
  1000, & 2\,\mathrm{cm}\leq x\leq3\,\mathrm{cm},
 \end{cases}
 &&\text{thin-to-thick},\\[2mm]
 \sigma_0(x)
 &=
 \begin{cases}
  1000, & 0\leq x<0.5\,\mathrm{cm},\\
  10,   & 0.5\,\mathrm{cm}\leq x\leq1.5\,\mathrm{cm},
 \end{cases}
 &&\text{thick-to-thin}.
 \end{aligned}
\]
The initial and boundary conditions,
material heat capacity, frequency discretization, angular quadrature,
and adaptive tolerances are those of
Section~\ref{sec:homogeneousMarshak}. The thin-to-thick case uses
\(600\times2\) uniform cells and is advanced to \(1\,\mathrm{ns}\); the
thick-to-thin case uses \(300\times2\) uniform cells and is advanced to
\(5\,\mathrm{ns}\). Both cases use a time step of
\(8.3389\times10^{-5}\,\mathrm{ns}\).

\begin{figure}[!htbp]
 \centering
 \includegraphics[width=\textwidth]{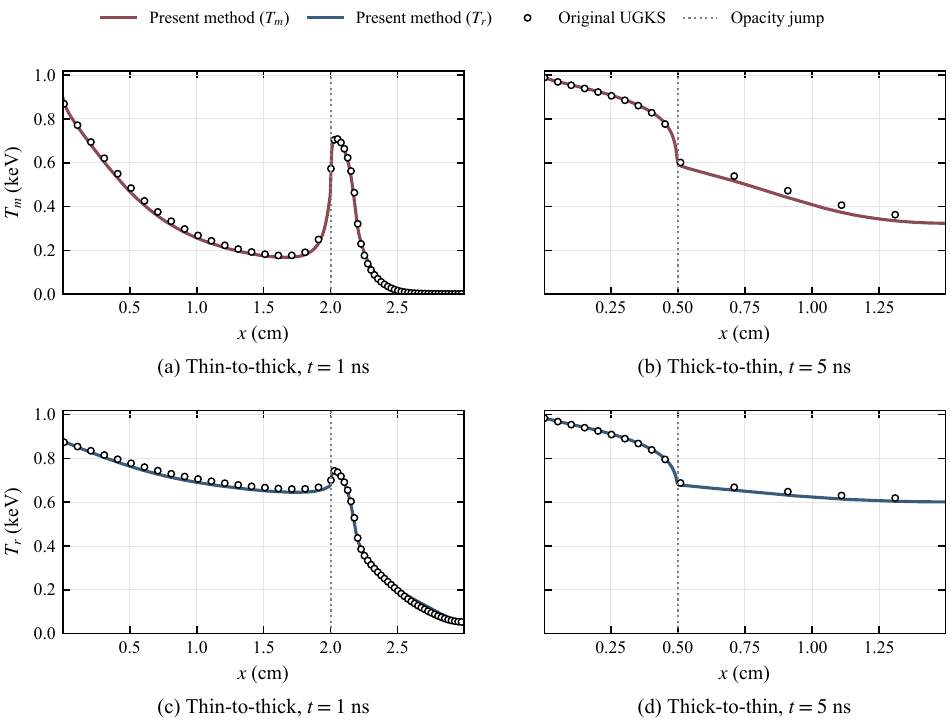}
 \caption{Heterogeneous Marshak wave: material temperature (a,b) and radiation
 temperature (c,d) in the thin-to-thick and thick-to-thin cases.
 Solid curves denote the present method; symbols denote the original UGKS
 results of Sun et al.~\cite{SunJiangXuLi2015}. Dotted vertical lines mark
 the opacity discontinuities.}
 \label{fig:heterogeneousMarshakProfiles}
\end{figure}

Figure~\ref{fig:heterogeneousMarshakProfiles} shows that the material and
radiation temperature results agree closely with the original UGKS results
in both cases.
They resolve the temperature variation across the material interface and
the radiation front in the downstream region.

\begin{figure}[!htbp]
 \centering
 \includegraphics[width=\textwidth]{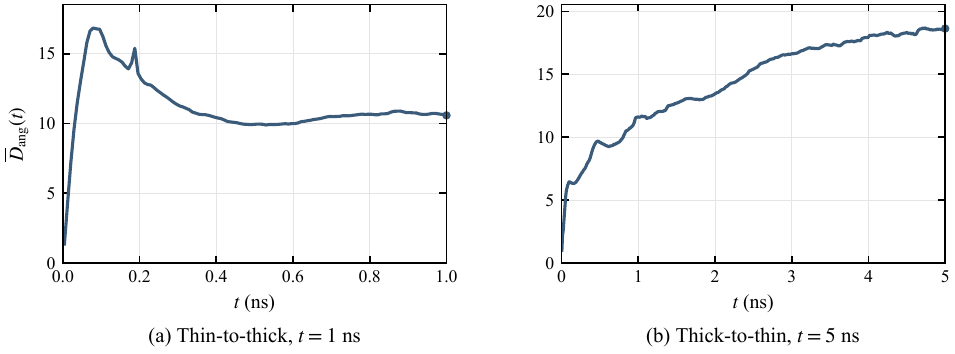}
 \caption{Mean retained DoFs for the heterogeneous Marshak wave.
 Circles mark the final values.}
 \label{fig:heterogeneousMarshakDofHistory}
\end{figure}

In the thin-to-thick case,
\(\overline D_{\mathrm{ang}}\) reaches \(16.85\) near
\(t=0.08\,\mathrm{ns}\), then decreases to \(10.58\), as shown in
Fig.~\ref{fig:heterogeneousMarshakDofHistory}.
In the thick-to-thin case, the mean number of retained DoFs reaches \(18.68\)
near \(t=4.73\,\mathrm{ns}\) and ends at \(18.64\).
The final retained fractions listed in Table~\ref{tab:heterogeneousMarshakResults}
are \(0.918\%\) and \(1.618\%\), respectively.

\begin{table}[!htbp]
 \centering
 \caption{Retained DoFs for the heterogeneous Marshak wave at
 the final time \(t_f\).}
 \label{tab:heterogeneousMarshakResults}
 \normalsize
 \renewcommand{\arraystretch}{1.18}
 \begin{tabular*}{0.86\textwidth}{@{\hspace{8pt}\extracolsep{\fill}}lcccc@{\hspace{8pt}}}
 \toprule
 Case & \(t_f\) (ns) & \(\overline D_{\mathrm{ang}}\) &
 \(f_{\mathrm{ang}}\) (\%) & \(p_{\max}\) \\
 \midrule
 Thin-to-thick & 1 & 10.58 & 0.918 & 11 \\
 Thick-to-thin & 5 & 18.64 & 1.618 & 12 \\
 \bottomrule
 \end{tabular*}
\end{table}

As summarized in Table~\ref{tab:heterogeneousMarshakMemory}, the speedups
are \(66.3\) and \(66.5\), and the memory savings exceed
\(98.8\%\) in both cases.

\begin{table}[!htbp]
 \centering
 \caption{Computational cost and memory for the heterogeneous Marshak wave.}
 \label{tab:heterogeneousMarshakMemory}
 \small
 \setlength{\tabcolsep}{4pt}
 \begin{tabular}{llrrrr}
 \toprule
 Case & Quantity & Original UGKS & Present method & Ratio & Reduction (\%) \\
 \midrule
 Thin-to-thick & Mean step time (s) & 1.626 & 0.02452 & 66.3 & --- \\
              & \(M\) (MiB) & 2545.18 & 29.42 & 86.52 & 98.84 \\
 \addlinespace
 Thick-to-thin & Mean step time (s) & 0.7978 & 0.01200 & 66.5 & --- \\
              & \(M\) (MiB) & 1276.82 & 15.22 & 83.91 & 98.81 \\
 \bottomrule
 \end{tabular}
\end{table}

\FloatBarrier
\subsection{Larsen problem}
\label{sec:larsenProblem}

The frequency-dependent Larsen problem~\cite{YangZhuLiuXu2025,Larsen1988GTA} is tested.
The physical domain is
\(\left[0,4\right]\times\left[0,0.04\right]\,\mathrm{cm}^{2}\). The opacity is
\[
 \sigma_{\nu}(x,T)
 =
 \gamma_0(x)\frac{1-\exp(-h_{\mathrm P}\nu/T)}
 {(h_{\mathrm P}\nu)^3},
 \qquad
 \gamma_0(x)
 =
 \begin{cases}
  1,    & 0\leq x\leq2\,\mathrm{cm},\\
  1000, & 2<x\leq3\,\mathrm{cm},\\
  1,    & 3<x\leq4\,\mathrm{cm},
 \end{cases}
\]
where \(\gamma_0\) is measured in
\(\mathrm{keV}^{3}\mathrm{cm}^{-1}\). Initially,
\(T_m=T_r=10^{-3}\,\mathrm{keV}\). The material energy is
\(U_m=C_vT_m\), with
\(C_v=0.05109\,\mathrm{GJ}\,\mathrm{cm}^{-3}\mathrm{keV}^{-1}\). Planck radiation at \(1\,\mathrm{keV}\) is injected through the left boundary
at \(x=0\). The right boundary is vacuum, and the transverse boundaries are reflecting.

The mesh has \(200\times2\) cells. The frequency interval corresponding to
photon energies \(\epsilon=h_{\mathrm P}\nu\in[10^{-2},10^{2}]\,\mathrm{keV}\)
is divided into 24 logarithmically spaced groups. The time step is
\(3.3356\times10^{-4}\,\mathrm{ns}\), and the solution is advanced to
\(t=0.9\,\mathrm{ns}\).
The angular tolerances are
\mbox{\(\varepsilon_{\mathrm{surv}}=10^{-2}\)} and
\mbox{\(\varepsilon_{\mathrm{flux}}=\varepsilon_{\mathrm{state}}=10^{-4}\)}.

\begin{figure}[!htbp]
 \centering
 \includegraphics[width=\textwidth]{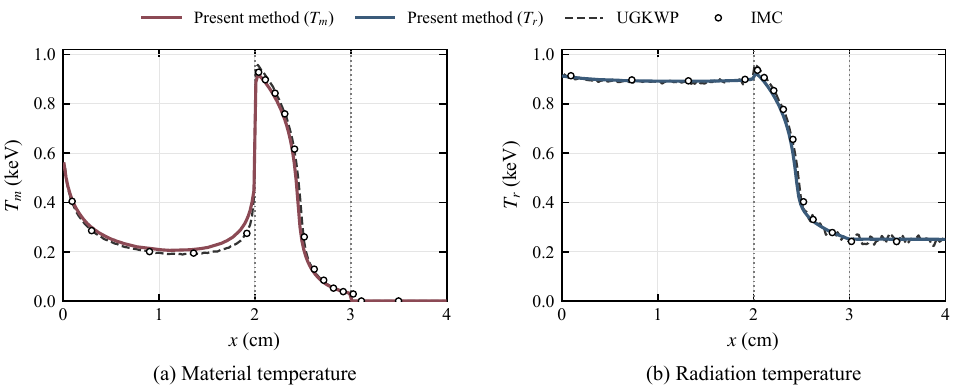}
 \caption{Larsen problem at \(t=0.9\,\mathrm{ns}\): (a) material temperature;
 (b) radiation temperature. Dotted vertical lines mark the opacity interfaces.}
 \label{fig:larsenTemperatureProfiles}
\end{figure}

The material and radiation temperature results agree closely with the IMC and
UGKWP results, as shown in Fig.~\ref{fig:larsenTemperatureProfiles}. The material
temperature has a sharp maximum at the left boundary of the high-opacity
interval \(2<x<3\,\mathrm{cm}\), and both temperatures decrease sharply
across this layer.

\begin{figure}[!htbp]
 \centering
 \includegraphics[width=\textwidth]{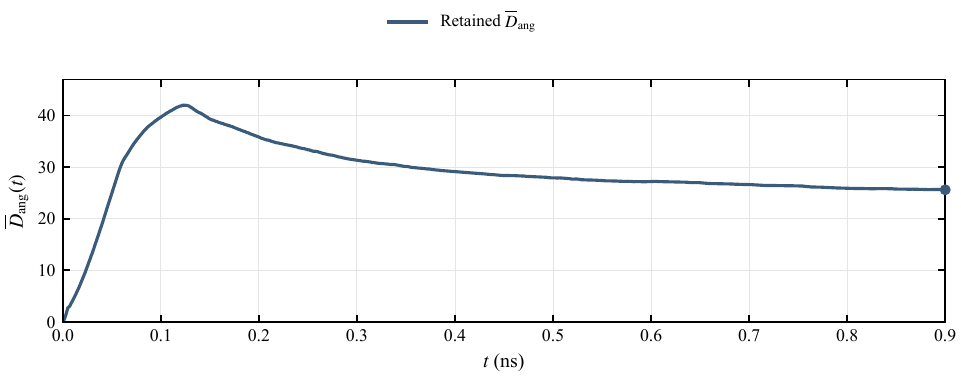}
 \caption{Mean retained DoFs for the Larsen
 problem. The circle marks the final mean retained DoFs.}
 \label{fig:larsenDofHistory}
\end{figure}

The mean number of retained DoFs reaches \(42.02\) at \(t=0.124\,\mathrm{ns}\) and
decreases to \(25.61\) at \(t=0.9\,\mathrm{ns}\), as shown in
Fig.~\ref{fig:larsenDofHistory}. The final retained DoFs are only
\(2.223\%\) of those in the original UGKS, as listed in
Table~\ref{tab:larsenAngularDimension}.

\begin{table}[!htbp]
 \centering
 \caption{Retained DoFs for the Larsen problem at \(t=0.9\,\mathrm{ns}\).}
 \label{tab:larsenAngularDimension}
 \normalsize
 \renewcommand{\arraystretch}{1.18}
 \begin{tabular*}{0.86\textwidth}{@{\hspace{8pt}\extracolsep{\fill}}ccc@{\hspace{8pt}}}
 \toprule
 \(\overline D_{\mathrm{ang}}\) & \(f_{\mathrm{ang}}\) (\%) & \(p_{\max}\) \\
 \midrule
 25.61 & 2.223 & 20 \\
 \bottomrule
 \end{tabular*}
\end{table}

The present method gives a speedup of \(46.46\) and reduces memory
by \(96.97\%\), as shown in Table~\ref{tab:larsenMemory}.

\begin{table}[!htbp]
 \centering
 \caption{Computational cost and memory for the Larsen problem.}
 \label{tab:larsenMemory}
 \small
 \setlength{\tabcolsep}{4pt}
 \begin{tabular}{lrrrr}
 \toprule
 Quantity & Original UGKS & Present method & Ratio & Reduction (\%) \\
 \midrule
 Mean step time (s) & 0.4472 & 0.009626 & 46.46 & --- \\
 \(M\) (MiB) & 854.03 & 25.84 & 33.05 & 96.97 \\
 \bottomrule
 \end{tabular}
\end{table}

\FloatBarrier
\subsection{Olson 1D problem}
\label{sec:olson1d}

The Olson carbon--hydrogen foam problem is tested in this section. The physical domain is
\([0,4.8]\times[0,0.075]\,\mathrm{cm}^{2}\). Let
\(\epsilon=h_{\mathrm P}\nu\), \(T_{\rm keV}=1\,\mathrm{keV}\), and
\(\theta=T/T_{\rm keV}\). The opacity is
\(\sigma_{\nu}=\varrho_m\kappa(\epsilon,T)\), where
\(\varrho_m=10^{-3}\,\mathrm{g}\,\mathrm{cm}^{-3}\) and
\begin{equation}
 \kappa(\epsilon,T)
 =
 \begin{cases}
  \min\{10^{7},10^{9}\theta^{2}\},
  & \epsilon\leq0.008,\\[1mm]
  \displaystyle
  \frac{3\times10^{6}(0.008/\epsilon)^{2}}
       {1+200\theta^{3/2}},
  & 0.008<\epsilon\leq0.3,\\[3mm]
  \displaystyle
  \frac{3\times10^{6}(0.008/\epsilon)^{2}(0.3/\epsilon)^{1/2}}
       {1+200\theta^{3/2}}
  +
  \frac{4\times10^{4}(0.3/\epsilon)^{5/2}}
       {1+8000\theta^{2}},
  & \epsilon>0.3.
 \end{cases}
 \label{eq:olson1dOpacity}
\end{equation}
Here \(\epsilon\) is measured in keV and \(\kappa\) in
\(\mathrm{cm}^{2}\,\mathrm{g}^{-1}\). The heat capacity is
\begin{equation}
 \begin{aligned}
  \frac{\dd U_m}{\dd T}
  &=aT_{\rm keV}^{3}H
  \left[1+\alpha+(T+T_\star)\frac{\partial\alpha}{\partial T}\right],\\
  \alpha
  &=\frac{e^{-T_\star/T}}{2}
  \left(\sqrt{1+4e^{T_\star/T}}-1\right),\\
  \frac{\partial\alpha}{\partial T}
  &=\frac{T_\star}{T^{2}}
  \left(\alpha-\frac{1}{\sqrt{1+4e^{T_\star/T}}}\right),
 \end{aligned}
 \label{eq:olson1dHeatCapacity}
\end{equation}
with \(T_\star=0.1T_{\rm keV}\), \(H=0.1\,\mathrm{cm}\), and
\(a=c=1\).

Initially, the radiation and material are in equilibrium at
\(0.01\,\mathrm{keV}\). All boundaries are reflecting. The isotropic volumetric
source, treated by operator splitting, is
\[
 Q_{\nu}(x,t)
 =
 \begin{cases}
  B_{\nu}(0.5\,\mathrm{keV})\exp(-693x^{3}), & 0\leq ct\leq2,\\
  0, & ct>2.
 \end{cases}
\]
The mesh contains \(128\times2\) cells. The frequency interval corresponding to
photon energies \(\epsilon=h_{\mathrm P}\nu\in[10^{-2},10^{2}]\,\mathrm{keV}\)
is divided into 24 logarithmically spaced groups. The time step is
\(c\Delta t=0.01875\,\mathrm{cm}\), and the solution is advanced to
\(ct=4\).
The angular tolerances are
\mbox{\(\varepsilon_{\mathrm{surv}}=10^{-2}\)} and
\mbox{\(\varepsilon_{\mathrm{flux}}=\varepsilon_{\mathrm{state}}=10^{-4}\)}.

\begin{figure}[!htbp]
 \centering
 \includegraphics[width=\textwidth]{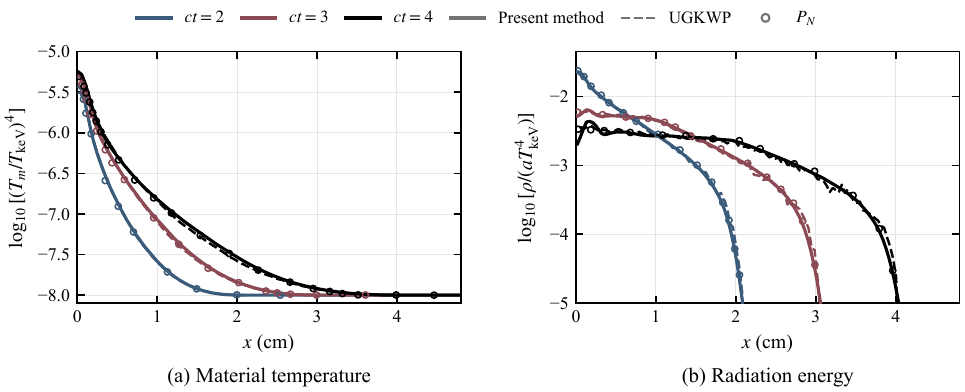}
 \caption{Olson 1D problem at \(ct=2,3,4\): (a) material
 temperature; (b) radiation energy. Solid curves denote the present method,
 dashed curves the UGKWP results, and symbols the \(P_N\) results of
 Olson~\cite{Olson2020MultigroupPN}. Both reference data sets are extracted
 from Fig.~7 of Yang et al.~\cite{YangZhuLiuXu2025}.}
 \label{fig:olson1dProfiles}
\end{figure}

The material temperature and radiation energy results at \(ct=2,3,4\)
agree closely with the UGKWP results of Yang et al.~\cite{YangZhuLiuXu2025}
and the \(P_N\) results of Olson~\cite{Olson2020MultigroupPN}, as shown in
Fig.~\ref{fig:olson1dProfiles}. The agreement persists after the source is
removed at \(ct=2\), as the radiation front propagates and broadens.

\begin{figure}[!htbp]
 \centering
 \includegraphics[width=\textwidth]{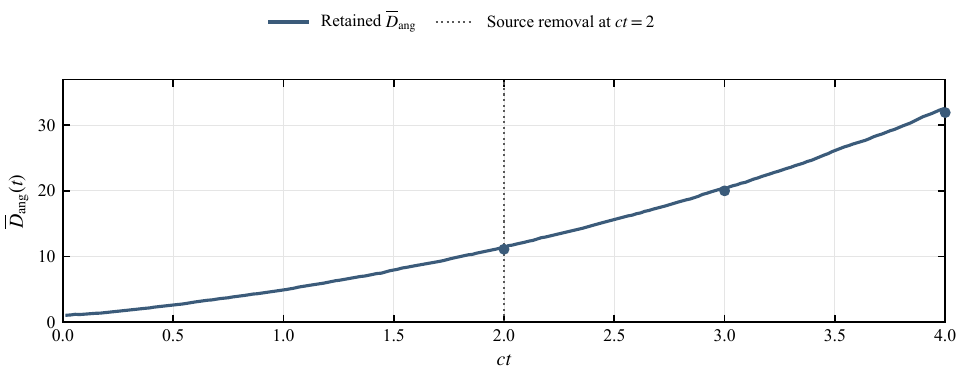}
 \caption{Mean retained DoFs for the Olson 1D
 problem. The dotted vertical line marks source removal; circles
 mark the mean retained DoFs at \(ct=2,3,4\).}
 \label{fig:olson1dDofHistory}
\end{figure}

As shown in Fig.~\ref{fig:olson1dDofHistory}, the mean number of retained
DoFs continues to increase after the source is removed. Between \(ct=2\) and \(ct=4\),
\(\overline D_{\mathrm{ang}}\) increases from \(11.09\) to \(31.94\), while
\(p_{\max}\) increases only from \(14\) to \(18\), as summarized in
Table~\ref{tab:olson1dAngularDimension}.

\begin{table}[!htbp]
 \centering
 \caption{Retained DoFs for the Olson 1D problem at \(ct=2,3,4\).}
 \label{tab:olson1dAngularDimension}
 \normalsize
 \renewcommand{\arraystretch}{1.18}
 \begin{tabular*}{0.86\textwidth}{@{\hspace{8pt}\extracolsep{\fill}}cccc@{\hspace{8pt}}}
 \toprule
 \(ct\) & \(\overline D_{\mathrm{ang}}\) & \(f_{\mathrm{ang}}\) (\%) & \(p_{\max}\) \\
 \midrule
 2 & 11.09 & 0.963 & 14 \\
 3 & 20.00 & 1.736 & 16 \\
 4 & 31.94 & 2.773 & 18 \\
 \bottomrule
 \end{tabular*}
\end{table}

As summarized in Table~\ref{tab:olson1dMemory}, the speedup is \(48.03\),
and the memory requirement is reduced by \(98.19\%\).

\begin{table}[!htbp]
 \centering
 \caption{Computational cost and memory for the Olson 1D problem.}
 \label{tab:olson1dMemory}
 \small
 \setlength{\tabcolsep}{4pt}
 \begin{tabular}{lrrrr}
 \toprule
 Quantity & Original UGKS & Present method & Ratio & Reduction (\%) \\
 \midrule
 Mean step time (s) & 0.3543 & 0.007377 & 48.03 & --- \\
 \(M\) (MiB) & 549.62 & 9.93 & 55.35 & 98.19 \\
 \bottomrule
 \end{tabular}
\end{table}

\FloatBarrier
\subsection{Two-dimensional Tophat problem}
\label{sec:tophat2d}

\begin{figure}[H]
 \centering
 \includegraphics[width=0.72\textwidth]{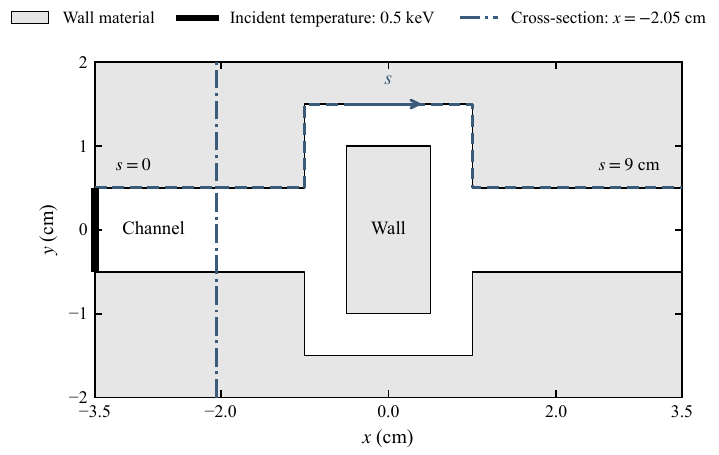}
 \caption{Geometry of the Tophat problem. The blackbody inflow is imposed at the
 left opening. The dashed upper-boundary path defines the arclength
 \(s\) in Fig.~\ref{fig:tophat2dProfiles}, from \(s=0\) at the left boundary
 to \(s=9\,\mathrm{cm}\) at the right boundary. The dash-dotted line marks
 \(x=-2.05\,\mathrm{cm}\), used in Fig.~\ref{fig:tophat2dCrossSection}.}
 \label{fig:tophat2dGeometry}
\end{figure}

In this section, we test the gray Tophat problem, which
combines an optically thin channel with optically thick walls and a central
obstacle. The physical domain is
\([-3.5,3.5]\times[-2,2]\,\mathrm{cm}^{2}\). The geometry and material
parameters follow Gorodetsky et al.~\cite{Gorodetsky2025TTTT}.
The channel is the union of the strip \(|y|<0.5\) and the rectangle
\(|x|<1,\ |y|<1.5\), excluding the obstacle
\(|x|<0.5,\ |y|<1\); coordinates are in cm.
The remaining region consists of wall material, as shown in
Fig.~\ref{fig:tophat2dGeometry}. The absorption coefficient is
\(0.2\,\mathrm{cm}^{-1}\) in the channel and
\(2000\,\mathrm{cm}^{-1}\) in the walls. The material energy is
\(U_m=C_vT_m\), with volumetric heat capacities
\(C_v=10^{-3}\) and \(1\,\mathrm{GJ}\,\mathrm{cm}^{-3}\mathrm{keV}^{-1}\),
respectively. The scattering coefficient is zero.

Initially, the material and radiation are in equilibrium at
\(T_m=T_r=0.05\,\mathrm{keV}\). A constant blackbody inflow at
\(0.5\,\mathrm{keV}\) is imposed at the left opening,
\(x=-3.5\,\mathrm{cm}\), \(|y|<0.5\,\mathrm{cm}\).
Extrapolation boundary conditions are imposed on the radiation intensity at
the remaining external boundaries.

We solve the gray radiation transport model on a mesh of \(140\times80\)
uniform cells with \(\Delta x=\Delta y=0.05\,\mathrm{cm}\). We use
\(c=29.9792458\,\mathrm{cm}\,\mathrm{ns}^{-1}\) and
\(a=0.01372\,\mathrm{GJ}\,\mathrm{cm}^{-3}\mathrm{keV}^{-4}\).
The nominal time step is
\(\Delta t=6.67128190\times10^{-4}\,\mathrm{ns}\), corresponding to
\(\mathrm{CFL}=0.8\), and the computation is advanced to
\(10\,\mathrm{ns}\). The angular tolerances are
\mbox{\(\varepsilon_{\mathrm{surv}}=10^{-2}\)} and
\mbox{\(\varepsilon_{\mathrm{flux}}=\varepsilon_{\mathrm{state}}=10^{-4}\)}.

The IMC calculation uses the same mesh, material parameters, and
initial and boundary conditions as the present method.
It uses \(\mathrm{CFL}=10\), corresponding to a time step of
\(8.33910238\times10^{-3}\,\mathrm{ns}\), and takes \(1200\) steps to reach
\(10\,\mathrm{ns}\). The particle number is set to \(6.4\times10^6\).

\begin{figure}[!htbp]
 \centering
 \includegraphics[width=\textwidth]{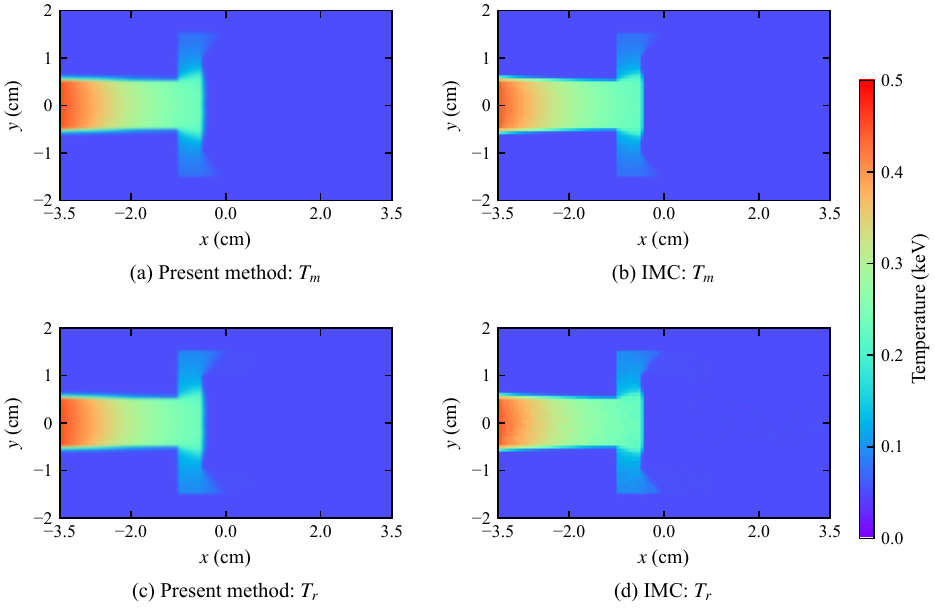}
 \caption{Tophat temperatures at \(t=10\,\mathrm{ns}\): material temperature
 (a,b) and radiation temperature (c,d) from the present method and IMC
 with \(\mathrm{CFL}=10\).}
 \label{fig:tophat2dFields}
\end{figure}

\begin{figure}[!htbp]
 \centering
 \includegraphics[width=\textwidth]{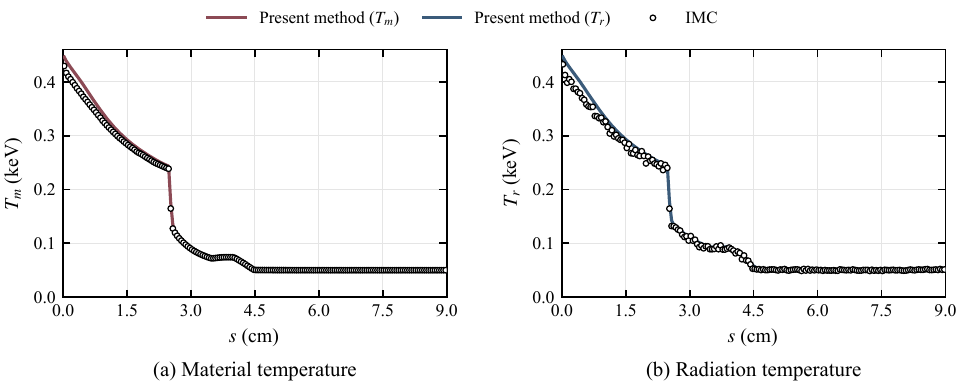}
 \caption{Tophat temperatures along the channel side of the upper wall at
 \(t=10\,\mathrm{ns}\): (a) material temperature; (b) radiation temperature.
 Here, \(s\) is the arclength along the path in Fig.~\ref{fig:tophat2dGeometry}.
 Solid curves denote the present method; symbols denote IMC with
 \(\mathrm{CFL}=10\).}
 \label{fig:tophat2dProfiles}
\end{figure}

\begin{figure}[!htbp]
 \centering
 \includegraphics[width=\textwidth]{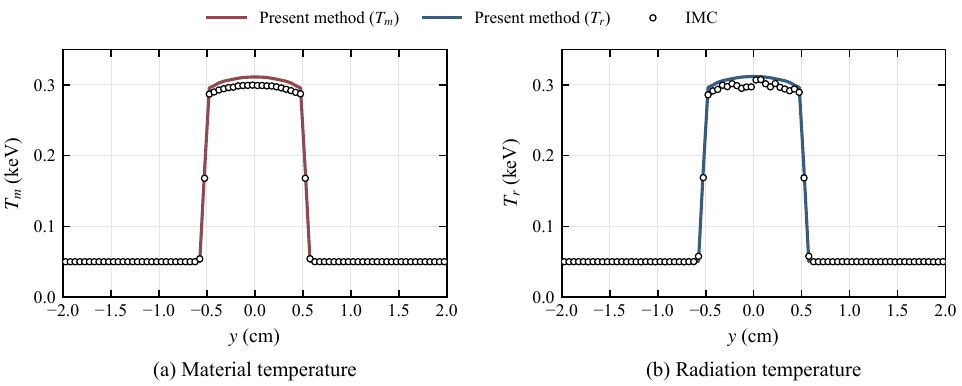}
 \caption{Tophat temperatures across the section at \(x=-2.05\,\mathrm{cm}\) and
 \(t=10\,\mathrm{ns}\): (a) material temperature; (b) radiation temperature.
 Solid curves denote the present method; symbols denote IMC with
 \(\mathrm{CFL}=10\).}
 \label{fig:tophat2dCrossSection}
\end{figure}

As shown in Fig.~\ref{fig:tophat2dFields}, the present method and IMC give
consistent material and radiation temperature fields, with a heated entrance
channel and a cold region behind the central obstacle. The temperature
distributions along the channel side of the upper wall and across the section
at \(x=-2.05\,\mathrm{cm}\) also agree well with IMC, as shown in
Figs.~\ref{fig:tophat2dProfiles} and \ref{fig:tophat2dCrossSection}.
Both methods show a sharp temperature decrease from the channel into the walls.

\begin{figure}[!htbp]
 \centering
 \includegraphics[width=\textwidth]{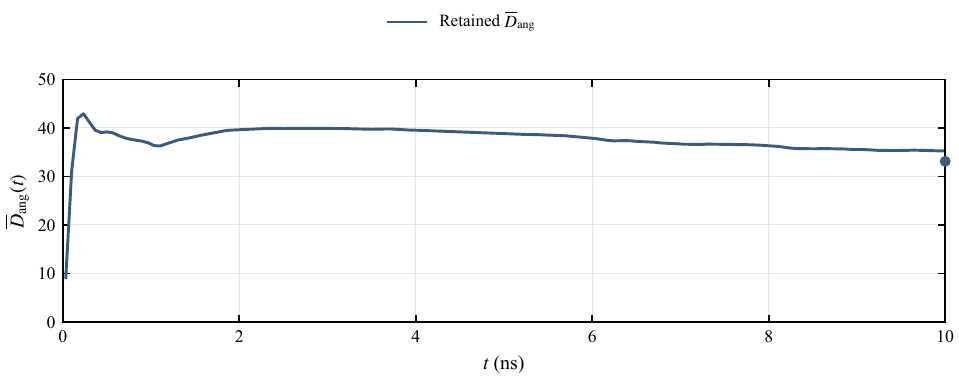}
 \caption{Mean retained DoFs for the Tophat problem.
 The circle marks the final value.}
 \label{fig:tophat2dDofHistory}
\end{figure}

Over all time steps, \(\overline D_{\mathrm{ang}}\) averages \(37.69\), or
\(3.27\%\) of the original UGKS value. It reaches a maximum of \(43.13\)
near \(t=0.22\,\mathrm{ns}\), corresponding to \(3.74\%\).
As shown in Fig.~\ref{fig:tophat2dDofHistory}, the mean retained DoFs decrease
gradually for \(t>3\,\mathrm{ns}\), reaching \(33.11\) at
\(t=10\,\mathrm{ns}\), when \(p_{\max}=20\).
Table~\ref{tab:tophat2dPerformance} summarizes the computational cost: the mean
time per step decreases from \(0.3624\) to \(0.01457\,\mathrm{s}\), giving a
speedup of \(24.87\). The memory requirement is reduced by \(95.39\%\).

\begin{table}[!htbp]
 \centering
 \caption{Computational cost and memory for the Tophat problem.}
 \label{tab:tophat2dPerformance}
 \small
 \setlength{\tabcolsep}{4pt}
 \begin{tabular}{lrrrr}
 \toprule
 Quantity & Original UGKS & Present method & Ratio & Reduction (\%) \\
 \midrule
 Mean step time (s) & 0.3624 & 0.01457 & 24.87 & --- \\
 \(M\) (MiB) & 511.70 & 23.59 & 21.69 & 95.39 \\
 \bottomrule
 \end{tabular}
\end{table}

Table~\ref{tab:tophat2dIMCTiming} compares the wall times of NCUGKS and IMC.
NCUGKS reaches \(t=10\,\mathrm{ns}\) in \(218.86\,\mathrm{s}\), compared with
\(6453.83\,\mathrm{s}\) for IMC, giving a speedup of \(29.49\).

\begin{table}[!htbp]
 \centering
 \caption{Time steps, wall times, and speedups relative to IMC for the Tophat problem at
 \(t=10\,\mathrm{ns}\).}
 \label{tab:tophat2dIMCTiming}
 \small
 \setlength{\tabcolsep}{8pt}
 \begin{tabular}{lrrr}
 \toprule
 Method & Steps & Wall time (s) & Speedup \\
 \midrule
 NCUGKS & 14990 & 218.86 & 29.49 \\
 IMC & 1200 & 6453.83 & 1.00 \\
 \bottomrule
 \end{tabular}
\end{table}

\FloatBarrier
\subsection{Olson 2D problem}
\label{sec:olson2d}

The two-dimensional Olson problem~\cite{YangZhuLiuXu2025,Olson2020MultigroupPN}
is tested in this section. The physical domain is
\([0,3.8]\times[0,3.8]\,\mathrm{cm}^{2}\). Three rectangular aluminum
inclusions are embedded in CH foam. The aluminum region is
\[
 \begin{aligned}
 \mathcal A={}&[0.5,1.5]\times[0.5,1.5]
 \;\cup\;[1.5,2.5]\times[1.5,2.5]\\
 &\cup\;[1.5,2.5]\times[0,0.5]
 \qquad (\mathrm{cm}^{2}).
 \end{aligned}
\]
The opacity and heat capacity of the foam are given by
Eqs.~\eqref{eq:olson1dOpacity} and \eqref{eq:olson1dHeatCapacity}. The same
heat-capacity relation is used for aluminum, with
\(T_\star=0.3\,\mathrm{keV}\) and \(H=0.5\,\mathrm{cm}\). The corresponding foam
parameters are \(0.1\,\mathrm{keV}\) and \(0.1\,\mathrm{cm}\).
As in the Olson 1D problem, \(a=c=1\).
The aluminum opacity is
\(\sigma_\nu=\varrho_m\kappa_{\rm Al}(\epsilon,T)\), where
\(\varrho_m=10^{-3}\,\mathrm{g}\,\mathrm{cm}^{-3}\) and
\[
\kappa_{\rm Al}(\epsilon,T)=
\begin{cases}
 \min\{10^{7},10^{8}\theta^2\}, & \epsilon\leq0.01,\\[1mm]
 \displaystyle
 \frac{10^{7}(0.01/\epsilon)^2}{1+20\theta^{3/2}},
 & 0.01<\epsilon\leq0.1,\\[3mm]
 \displaystyle
 \frac{10^{7}(0.01/\epsilon)^2}{1+20\theta^{3/2}}
 +\frac{10^{6}(0.1/\epsilon)^2}{1+200\theta^2},
 & 0.1<\epsilon\leq1.5,\\[3mm]
 \displaystyle
 \frac{10^{7}(0.01/\epsilon)^2(1.5/\epsilon)^{1/2}}
      {1+20\theta^{3/2}}
 +\frac{10^{5}(1.5/\epsilon)^{5/2}}{1+1000\theta^2},
 & \epsilon>1.5.
\end{cases}
\]
Here \(\epsilon\) is measured in keV and \(\kappa_{\rm Al}\) in
\(\mathrm{cm}^{2}\,\mathrm{g}^{-1}\).

Initially, the radiation and material are in equilibrium at
\(0.01\,\mathrm{keV}\). The isotropic volumetric source, treated
by operator splitting, is
\[
 Q_{\nu}(x,y,t)
 =
 \begin{cases}
 B_{\nu}(0.5\,\mathrm{keV})
 \exp[-18.7(x^2+y^2)^{3/2}], & 0\leq ct\leq3,\\
 0, & ct>3.
 \end{cases}
\]
All boundaries are reflecting. The mesh has \(76\times76\) cells. The
frequency interval corresponding to photon energies
\(\epsilon=h_{\mathrm P}\nu\in[10^{-2},10^{2}]\,\mathrm{keV}\)
is divided into 24 logarithmically spaced groups. The time step is
\(c\Delta t=2\times10^{-2}\,\mathrm{cm}\), and the solution is
advanced to \(ct=3\). The angular tolerances are
\mbox{\(\varepsilon_{\mathrm{surv}}=10^{-2}\)} and
\mbox{\(\varepsilon_{\mathrm{flux}}=\varepsilon_{\mathrm{state}}=10^{-5}\)}.

\begin{figure}[!htbp]
 \centering
 \includegraphics[width=\textwidth]{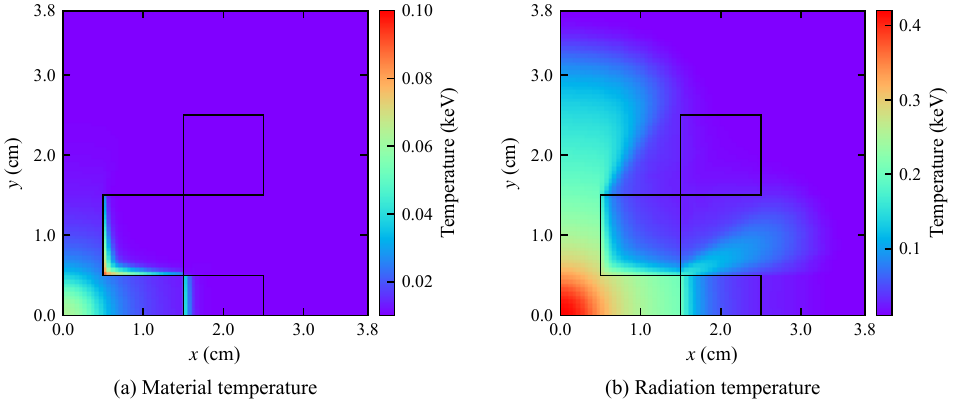}
 \caption{Olson 2D problem on the \(76\times76\) mesh at \(ct=3\): (a) material temperature;
 (b) radiation temperature. Black lines mark the aluminum interfaces.}
 \label{fig:olson2dFields}
\end{figure}

The material and radiation temperature results in Fig.~\ref{fig:olson2dFields}
show low-temperature regions behind the aluminum inclusions at \(ct=3\).

\begin{figure}[!htbp]
 \centering
 \includegraphics[width=\textwidth]{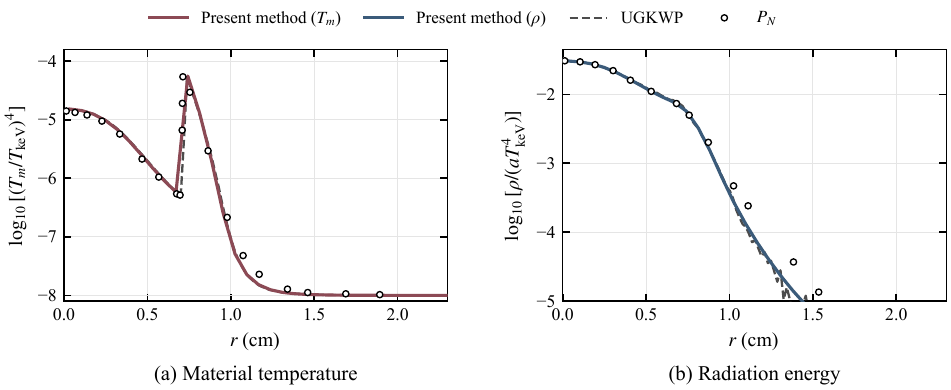}
 \caption{Olson 2D results along \(y=x\) at \(ct=3\): (a) material
 temperature; (b) radiation energy. The coordinate is
 \(r=\sqrt{x^2+y^2}=\sqrt{2}\,x\). Solid curves denote the present method,
 dashed curves the UGKWP results, and symbols the \(P_N\) results of
 Olson~\cite{Olson2020MultigroupPN}. Both reference data sets are extracted
 from Fig.~9 of Yang et al.~\cite{YangZhuLiuXu2025}.}
 \label{fig:olson2dProfiles}
\end{figure}

The material temperature and radiation energy results along \(y=x\)
agree well with the UGKWP results of Yang et al.~\cite{YangZhuLiuXu2025}
and the \(P_N\) results of Olson~\cite{Olson2020MultigroupPN}, as shown in
Fig.~\ref{fig:olson2dProfiles}.

\begin{figure}[!htbp]
 \centering
 \includegraphics[width=\textwidth]{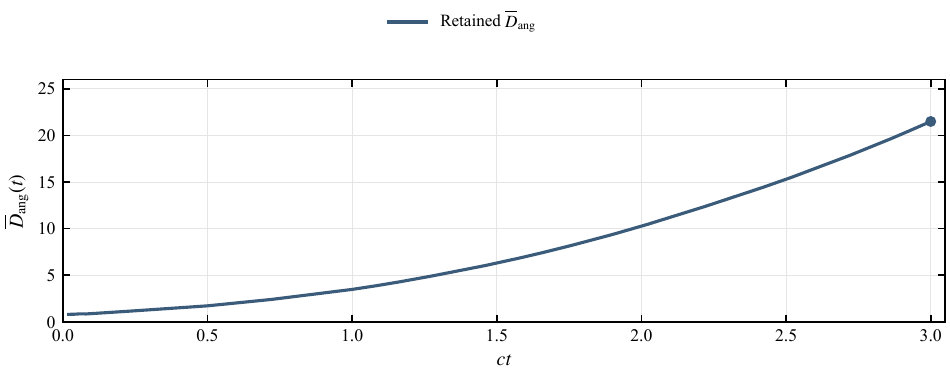}
 \caption{Mean retained DoFs for the Olson 2D problem.
 The circle marks the final value.}
 \label{fig:olson2dDofHistory}
\end{figure}

As shown in Fig.~\ref{fig:olson2dDofHistory}, the mean number of retained
DoFs generally increases with time. At \(ct=3\),
\(\overline D_{\mathrm{ang}}=21.49\), \(f_{\mathrm{ang}}=1.866\%\), and
\(p_{\max}=19\).

Mean step times are averaged over the complete 150-step simulations.
The mean step time decreases from \(6.272\) to \(0.1187\,\mathrm{s}\), giving
a speedup of \(52.84\), as summarized in Table~\ref{tab:olson2dPerformance}.
The memory requirement is \(161.93\,\mathrm{MiB}\), compared with
\(4873.50\,\mathrm{MiB}\) for the original UGKS, a saving of \(96.68\%\).

\begin{table}[!htbp]
 \centering
 \caption{Computational cost and memory for the Olson 2D problem.}
 \label{tab:olson2dPerformance}
 \small
 \setlength{\tabcolsep}{4pt}
 \begin{tabular}{lrrrr}
 \toprule
 Quantity & Original UGKS & Present method & Ratio & Reduction (\%) \\
 \midrule
 Mean step time (s) & 6.272 & 0.1187 & 52.84 & --- \\
 \(M\) (MiB) & 4873.50 & 161.93 & 30.10 & 96.68 \\
 \bottomrule
 \end{tabular}
\end{table}

\FloatBarrier

\section{Conclusions}
\label{sec:conclusion}

\begingroup
\hyphenpenalty=10000
\emergencystretch=1em
This work develops NCUGKS for frequency dependent thermal
radiative transfer in which the angular intensity is represented in an
orthonormal basis of the discrete angular space. The method retains the UGKS
time dependent interface evolution and multiscale treatment of nonequilibrium radiation.

The basis coefficients to be evolved are selected using the exponentially decayed
contribution from the previous time step and the requirements of interface transport.
After the finite volume update, higher-degree coefficients are discarded
according to the prescribed relative tail tolerance. The flux truncation error
is bounded by the norms of the discarded basis coefficients of the reconstructed
intensity and its spatial derivatives. In the optically thick limit, each
quadrant retains only its constant coefficient and the flux truncation error
tends to zero. The candidate and interface space dimensions then remain bounded
independently of the number of angular quadrature nodes.

The method was evaluated on one and two dimensional tests spanning different
optical regimes. The solutions agree closely with published reference solutions and results obtained
with other methods. Relative to the full discrete ordinate UGKS, memory
is reduced by factors from \(21.7\) to \(154.0\), while the measured speedups
range from \(24.87\) to \(257.9\), demonstrating that NCUGKS accurately captures
nonequilibrium radiation transport with substantially fewer angular DoFs
that adapt to local optical thickness.

\endgroup

\section*{Acknowledgements}
The current research is supported by the National Natural Science Foundation of China (92371107, 12302378), the Funding of National Key Laboratory of Computational Physics, the National Key R\&D Program of China (Grant No. 2022YFA1004500), the Natural Science Basic Research Plan in Shaanxi Province of China (No. 2025SYS-SYSZD-070), and the Hong Kong Research Grant Council (16208324).

\begingroup
\hyphenpenalty=10000
\bibliographystyle{elsarticle-num}
\bibliography{main}
\endgroup

\end{document}